\documentclass[11pt,danish,a4paper,floatfix]{article}
\pdfoutput=1
\usepackage{jcappub}

\usepackage{bm}
\usepackage{amsfonts}
\usepackage{subfigure}
\usepackage{booktabs}
\newcommand{\ra}[1]{\renewcommand{\arraystretch}{#1}}

\newcommand{\be}{\begin{equation}}
\newcommand{\ee}{\end{equation}}
\newcommand{\bea}{\begin{eqnarray}}
\newcommand{\eea}{\end{eqnarray}}

\graphicspath{{plots/}}
\begin{document}


\title{Updated results on neutrino mass and mass hierarchy from cosmology with Planck 2018 likelihoods}

\author[a,b]{Shouvik Roy Choudhury,}
\author[c]{Steen Hannestad}
\affiliation[a]{Department of Physics, Indian Institute of Technology Bombay,\\Main Gate Road, Powai, Mumbai 400076, India}

\affiliation[b]{Harish-Chandra Research Institute, HBNI,\\Chhatnag Road, Jhunsi, Allahabad 211019, India}

\affiliation[c]{Department of Physics and Astronomy, Aarhus University,\\
Ny Munkegade 120, DK-8000 Aarhus C, Denmark}

\emailAdd{shouvikroychoudhury@gmail.com, sth@phys.au.dk}

\abstract{In this work we update the bounds on $\sum m_{\nu}$ from latest publicly available cosmological data and likelihoods using Bayesian analysis, while explicitly considering particular neutrino mass hierarchies. In the minimal $\Lambda\textrm{CDM}+\sum m_{\nu}$ model with most recent CMB data from Planck 2018 TT,TE,EE, lowE, and lensing; and BAO data from BOSS DR12, MGS, and 6dFGS, we find that at 95\% C.L. the bounds are: $\sum m_{\nu}<0.12$ eV (degenerate), $\sum m_{\nu}<0.15$ eV (normal), $\sum m_{\nu}<0.17$ eV (inverted). The bounds vary across the different mass orderings due to different priors on $\sum m_{\nu}$. Also, we find that the normal hierarchy is very mildly preferred relative to the inverted, using both minimum $\chi^2$ values and Bayesian Evidence ratios. In this paper we also provide bounds on $\sum m_{\nu}$ considering different hierarchies in various extended cosmological models: $\Lambda\textrm{CDM}+\sum m_{\nu}+r$, $w\textrm{CDM}+\sum m_{\nu}$, $w_0 w_a \textrm{CDM}+\sum m_{\nu}$, $w_0 w_a \textrm{CDM}+\sum m_{\nu}$ with $w(z)\geq -1$, $\Lambda \textrm{CDM} + \sum m_{\nu} + \Omega_k$, and $\Lambda \textrm{CDM} + \sum m_{\nu} + A_{\textrm{Lens}}$. We do not find any strong evidence of normal hierarchy over inverted hierarchy in the extended models either.}

\maketitle


\section{Introduction} \label{sec1}
Earth based neutrino oscillation experiments \cite{Abe:2013hdq, Ahn:2012nd, Abe:2012tg, An:2012eh, Araki:2004mb, Adamson:2008zt, Fukuda:1998mi, Ahmad:2002jz} have confirmed that neutrinos are massive, which is the first known departure from the Standard Model of particle physics where neutrinos are considered to be massless. The 3 neutrino flavor states ($\nu_e$, $\nu_{\mu}$, $\nu_{\tau}$) are quantum superpositions of the 3 mass eigenstates ($\nu_i$, with respective distinct masses, $m_i$ for $i=1,2,3$).  However, because oscillation experiments use ultra relativistic neutrinos they are only sensitive to the mass-squared splittings ($\Delta m_{ij}^2 = m_i^2 - m_j^2$) and not the absolute masses, thus keeping the mass of the lightest neutrino unbounded. On the other hand, while magnitudes of $\Delta m_{21}^2$ and $\Delta m_{31}^2$ are known to considerable accuracy from the current oscillation data, sign of $\Delta m_{31}^2$ is unknown. This leads to two possible hierarchies of neutrino masses: $m_1<m_2\ll m_3$ (normal hierarchy or NH) and $m_3\ll m_1<m_2$ (inverted hierarchy or IH) depending on whether $\Delta m_{31}^2$ is positive or negative, respectively. As per the latest NuFit 4.0 \cite{Esteban:2018azc} global analysis of oscillations data, the values of the mass-squared splittings (in units of eV$^{2}$) are (limits are given at 1$\sigma$):
\begin{equation} \label{eq1}
 \Delta m_{21}^2 = 7.39^{+0.21}_{-0.20}\times 10^{-5};~~ \Delta m_{31}^2= 2.525^{+0.033}_{-0.032}\times 10^{-3} (\textrm{NH});~~\Delta m_{32}^2= - 2.512^{+0.034}_{-0.032}\times 10^{-3} (\textrm{IH}).   
\end{equation}
Here, the value for $\Delta m_{21}^2$ is applicable to both NH and IH, while the other values are for the particular hierarchies as mentioned in the brackets. It is to be noted that for IH, $\Delta m_{32}^2$ is provided (whose value is negative) instead of $\Delta m_{31}^2$, but since $\Delta m_{21}^2 \ll \Delta m_{32}^2$, sign of $\Delta m_{31}^2$ for IH is also negative (since $\Delta m_{31}^2 = \Delta m_{32}^2 + \Delta m_{21}^2$). See \cite{Forero:2014bxa, Gonzalez-Garcia:2014bfa, Esteban:2016qun, Capozzi:2016rtj, Capozzi:2017ipn, Caldwell:2017mqu} for results from other global analyses.

A solution to the neutrino mass hierarchy problem may come from cosmology, which currently provides the strongest bounds on the absolute neutrino mass scale, defined as the sum of the three active neutrino masses, 
\begin{equation} \label{eq2}
\sum m_{\nu} = m_1 + m_2 + m_3.
\end{equation}
As far as known physics goes, at temperatures $T \gg {\rm MeV}$, neutrinos remain in equilibrium with the primordial plasma through the standard model weak interactions. At around $T\sim {\rm MeV}$ neutrinos decouple from the plasma and start free streaming. When neutrinos are relativistic ($T \gg m_{\nu}$) they contribute to the radiation energy density. This continues until much later when they turn non-relativistic at temperatures $T \sim m_{\nu}$, and then they contribute to the matter energy density. Effects of massive neutrinos on cosmological observables have been extensively studied in the literature \cite{Lesgourgues:2006nd, Wong:2011ip, Lesgourgues:2012uu, Abazajian:2013oma, Lesgourgues:2014zoa, Archidiacono:2016lnv, Lattanzi:2017ubx} and these effects help in constraining the sum of neutrino masses. If we consider neutrinos with masses $\ll 1$ eV, at the time of photon decoupling they are still relativistic, and their mass has very limited effect on the photon perturbations and their evolution. Hence, for the primary CMB signal, the effect of small neutrino masses can only be seen through the background evolution, and secondary anisotropies like Integrated Sachs-Wolfe (ISW) effect, and these too can be compensated partially by varying other free parameters in the $\Lambda$CDM model. Thus, strong bounds on $\sum m_{\nu}$ cannot be obtained with CMB power spectrum data alone. On the other hand, at late times, neutrinos affect the evolution of matter perturbations to a large extent. Due to the free-streaming effect of neutrinos, i.e. large thermal velocities, neutrinos do not cluster on small length scales, and this causes increasing suppression of small scale matter power spectrum with increasing fraction of neutrino energy density with respect to the total matter density \cite{Lesgourgues:2014zoa}. Thus if we augment CMB anisotropy data with data coming Large Scale Structure (LSS), CMB lensing (which is the weak lensing effect on CMB photons due to LSS) measurements etc, strong bounds on $\sum m_{\nu}$ can be obtained. Even so, currently it is only possible to get an upper bound on $\sum m_{\nu}$ from cosmological data alone. 

Let us define the mass of the lightest neutrino mass eigenstate to be $m_0$. For normal hierarchy, $m_0 = m_1$, whereas for inverted hierarchy, $m_0 = m_3$. In terms of $m_0$, the sum of the neutrino masses can be defined as,
\begin{equation}\label{eq3}
\sum m_{\nu} = m_0 + \sqrt{\Delta m_{21}^2 + m_0^2} + \sqrt{\Delta m_{31}^2+ m_0^2} ~~~~~~~~~~~~~~~~~~~~~~~~(\textrm{NH}),
\end{equation} 
and
\begin{equation} \label{eq4}
\sum m_{\nu} = m_0 + \sqrt{|\Delta m_{32}^2| + m_0^2} + \sqrt{|\Delta m_{32}^2|- \Delta m_{21}^2 + m_0^2} ~~~~~~~~~~~~~(\textrm{IH}).
\end{equation}
Putting $m_0 = 0$, one can obtain the minimum neutrino mass sums allowed by the two possible hierarchies, and these are $\sum m_{\nu} = 0.05885^{+0.00045}_{-0.00044}$ eV (1$\sigma$)(NH) and $\sum m_{\nu} = 0.100^{+0.00070}_{-0.00067}$ eV (1$\sigma$)(IH). Assuming that the normal hierarchy is the true one, the way cosmology can help is by constraining the $\sum m_{\nu}$ below the minimum mass required by inverted hierarchy with reasonable statistical significance, e.g., a determination of $\sum m_{\nu} = 0.058\pm 0.008$ eV (1$\sigma$) will exclude inverted hierarchy at 5$\sigma$ and also provide cosmological evidence for non-zero neutrino masses at 7.25 $\sigma$. Currently, the most recent and strongest bounds on $\sum m_{\nu}$ in the minimal $\Lambda$CDM+$\sum m_{\nu}$ model are around $\sum m_{\nu} <0.12$ eV (95\% C.L.) \cite{Choudhury:2018byy, Aghanim:2018eyx} with CMB and BAO data, whereas some other studies had reported bounds around $\sum m_{\nu} < 0.15$ eV (95\%) or better \cite{Vagnozzi:2017ovm, Palanque-Delabrouille:2015pga,DiValentino:2015wba,Cuesta:2015iho,Huang:2015wrx,Moresco:2016nqq,Giusarma:2016phn,Couchot:2017pvz,Caldwell:2017mqu,Doux:2017tsv,Wang:2017htc,Chen:2017ayg,Upadhye:2017hdl,Salvati:2017rsn,Nunes:2017xon,Zennaro:2017qnp,Wang:2018lun, Choudhury:2018adz, Giusarma:2018jei,Loureiro:2018pdz}. The strongest bounds are very close to the minimum $\sum m_{\nu}$ required for inverted hierarchy and thus it seems that inverted hierarchy is starting to get under pressure from cosmological data. These bounds are obtained with the assumption that all the three neutrino masses are equal ($m_i = \sum m_{\nu}/3$ for $i=1,2,3$), an approximation we denote as degenerate hierarchy (DH). There are also studies covering interesting aspects of measurement of neutrino hierarchy from cosmology \cite{Hannestad:2016fog, Vagnozzi:2017ovm,Xu:2016ddc,Gerbino:2016ehw,Simpson:2017qvj,Schwetz:2017fey,Long:2017dru,Gariazzo:2018pei, Heavens:2018adv,deSalas:2018bym,Loureiro:2018pdz,Mahony:2019fyb}. The current cosmological data, however, is not yet sensitive enough to make the distinction between the two hierarchies on a level that can be considered statistically significant, but there is a small preference for normal hierarchy \cite{Hannestad:2016fog, Vagnozzi:2017ovm}. It is to be noted that the bounds depend on the underlying cosmological model, and any extensions to the minimal $\Lambda\textrm{CDM}+\sum m_{\nu}$ model will usually lead to a more relaxed bound on $\sum m_{\nu}$ \cite{Choudhury:2018byy, Vagnozzi:2017ovm, Wang:2017htc, Chen:2017ayg,Hannestad:2005gj,Joudaki:2012fx,Yang:2017amu,Lorenz:2017fgo,Sutherland:2018ghu,Sahlen:2018cku,DiValentino:2017zyq}. However, it can happen that the neutrino mass bound improves when the extension to $\Lambda$CDM is done in such a way that the allowed parameter space of the new parameters prefers neutrino masses which are smaller than what we get in $\Lambda$CDM. In fact this is the case when one incorporates dynamical dark energy in the cosmological model but restricts the parameter space to non-phantom dark energy only, i.e. neutrino mass bounds in a cosmology with non-phantom dynamical dark energy are stronger than that in $\Lambda\textrm{CDM}$ \cite{Choudhury:2018byy, Choudhury:2018adz, Vagnozzi:2018jhn}. Another such model where neutrino mass bounds are stronger than that in $\Lambda\textrm{CDM}$ is holographic dark energy (HDE) \cite{Zhang:2015uhk,Wang:2016tsz}. 

From now on, we shall use the abbreviations DH, NH, and IH for degenerate, normal and inverted hierarchies respectively. 
While the degenerate hierarchy approximation has been predominant in the literature, and makes sense when neutrino masses are relatively large compared to the square-root of the mass-squared splittings (i.e. $m_i \gg \sqrt{\Delta m_{ij}^2}$), the cosmological neutrino mass bound is becoming strong enough that it should be replaced by a treatment using either the normal or the inverted hierarchy.
Hence, in this paper we have updated the bounds on the $\sum m_{\nu}$ while explicitly considering three different hierarchies (degenerate, normal and inverted), using latest datasets and likelihoods that are publicly available, for the minimal $\Lambda\textrm{CDM} + \sum m_{\nu}$ and some of its extensions. Except in the case of extension with the tensor-to-scalar ratio ($r$) parameter, all the other extensions studied in this paper includes new parameters which are considerably correlated with the sum of neutrino masses in the datasets considered. Details of the models are given in the next section. The neutrino mass bounds are supposed to relax in most of the extended models, and the difference between the upper limits obtained for the three hierarchies is supposed to diminish as the individual masses become much larger than the square root of mass-squared splittings. But still, our motivation to study these extended models is to see whether the latest datasets can make a difference. 

To implement the normal and inverted hierarchy, we use the mean values of the mass squared splittings given in eq.(\ref{eq1}) along with the lightest neutrino mass $m_0$ to define $m_1$, $m_2$, and $m_3$, and use $m_0$ as a free parameter and $\sum m_{\nu}$ as a derived parameter. We ignore the errors in the measurement of the mass-squared splittings from oscillations data since they are small compared to the mean values and incorporating them would have a very small effect on the bounds on $\sum m_{\nu}$. For degenerate hierarchy, we simply have $\sum m_{\nu} = 3 m_0$. 

It is to be understood here that the physical parameter cosmological data is primarily sensitive to $\sum m_{\nu}$ and not the individual neutrino masses. Even if we consider very optimistic future cosmological data, same value of $\sum m_{\nu}$ but different neutrino mass hierarchies lead to changes which are smaller than the expected observational errors and modeling uncertainties, i.e. it will be impossible to differentiate individual neutrino masses even with future data \cite{Archidiacono:2020dvx}. Thus, while analysing cosmological data, it makes sense to vary $\sum m_{\nu}$ as the cosmological parameter. However, same is not true for other experiments such as neutrino oscillations, kinematic measurements from beta decay etc and thus using the lightest mass $m_0$ instead of $\sum m_{\nu}$ allows one to be more general in their approach, and leaves the possibility of incorporating non-cosmological datasets for further analysis \cite{Gerbino:2016ehw} open. If one uses $\sum m_{\nu}$ as the cosmological parameter, a good way to incorporate DH, NH, and IH is to use the following non-informative flat priors: $\sum m_{\nu} \geq 0$ (DH), $\sum m_{\nu} \geq 0.06$ (NH), $\sum m_{\nu} \geq 0.10$ eV (IH), with suitable upper limits. Due to eq. (\ref{eq3}) and (\ref{eq4}), if one uses a flat prior on $m_0$ instead of on $\sum m_{\nu}$, the effective prior on $\sum m_{\nu}$ is not guaranteed to be  for the NH and IH case, and requires further inspection. However, as shown in figure 6 of Ref \cite{Gerbino:2016ehw}, a flat prior on the lightest mass leads to an almost flat prior on $\sum m_{\nu}$, except that the prior probability rises only slightly close to the lowest possible $\sum m_{\nu}$ values in each hierarchical scenario. Thus it is expected that the bounds on $\sum m_{\nu}$ coming from the NH and IH scenarios with a flat prior on $m_0$ will be slightly stronger than what we can get by using a flat prior on $\sum m_{\nu}$. As we shall see that the differences in the bounds on $\sum m_{\nu}$ obtained using flat prior on $m_0$ are very close to the ones with a flat prior on $\sum m_{\nu}$ and thus using $m_0$ as a varying parameter is well-motivated.    

Another important point is that current cosmological datasets are only able to provide an upper bound to the $\sum m_{\nu}$. This bound changes when one changes the neutrino mass hierarchy model from DH to NH to IH. This happens due to volume effects, i.e. change in the prior on $\sum m_{\nu}$. However, in case future cosmological data is able to detect $\sum m_{\nu}$, it has been shown in \cite{DiValentino:2016foa}, that irrespective of the correct neutrino mass model (i.e. NH or IH), using the unrealistic DH approximation (with the prior $\sum m_{\nu} \geq 0$)  will actually lead to the recovery of the correct values of $\sum m_{\nu}$ with only a small reconstruction bias due to the assumption of wrong fitting model. A more recent study \cite{Archidiacono:2020dvx} also made a similar conclusion. Thus, the degenerate hierarchy approximation, while unrealistic, is still useful in cosmology. However, while using DH, we must be wary of any reconstruction bias in the recovered values of $\sum m_{\nu}$ in case it is detected from future data.     

We would like to point it out here that the choice for a proper prior on neutrino masses has been a topic of intense debate among the concerned researchers, especially because different prior choices can lead to very different bounds on $\sum m_{\nu}$ and very different inferences on the hierarchy issue. See \cite{Hannestad:2016fog,Gerbino:2016ehw,Vagnozzi:2017ovm,Long:2017dru,Gariazzo:2018pei,Heavens:2018adv,Handley:2018gel,Simpson:2017qvj,Schwetz:2017fey} for discussions on this topic.  

Among datasets, for CMB anisotropies we use the most recently released Planck 2018 likelihoods for the data on temperature, E mode polarisation and their cross-correlation. Other than Planck CMB anisotropies, we use latest data from measurements of Planck lensing, CMB B mode, BAO, and SNe Ia luminosity distance. 
   
This paper is structured as follows. In section \ref{sec2} we provide brief details about the cosmological models and datasets used in this paper. In section \ref{sec3} we provide and explain the results of our MCMC analyses. In section \ref{sec5} we also include results from additional analyses using nested sampling and provide a comparison of bayesian evidence between NH and IH, and quantify the evidence against IH closely following \cite{Hannestad:2016fog}. In section \ref{sec4} we conclude.



\section{Methodology: Models and datasets} \label{sec2} 

\subsection{Models}\label{sec2.1}
In this work we have performed our analyses using a variety of different cosmological models which we shall describe below. Note that when we label models we use the term $\sum m_\nu$ to refer to the neutrino mass. This is done in order to conform to the standard labelling used in cosmological parameter analyses. In fact all our runs use a flat prior on $m_0$, the mass of the lightest mass state, rather than a flat prior on $\sum m_\nu$.
In total we have investigated the following 7 sets of cosmological models:

\begin{itemize}
\item i) The minimal $\Lambda\textrm{CDM}+\sum m_{\nu}$ model: Below we list the vector of varying parameters in this model.
\begin{equation}
\theta \equiv \left[\omega_c, \omega_b, \Theta_{\textrm{s}}, \tau, n_{\textrm{s}}, \textrm{ln}[10^{10} A_{\textrm{s}}], m_0\right].
\end{equation}
Here the first six parameters correspond to the $\Lambda\textrm{CDM}$ model. $\omega_c = \Omega_c h^2$ and $\omega_b = \Omega_b h^2$ are the present cold dark matter and baryon energy densities respectively. $\Theta_{\textrm{s}}$ is the ratio between sound horizon $r_s$ and angular diameter distance $D_{\textrm{A}}$ at the time of photon decoupling. $\tau$ is the optical depth to re-ionization of the universe at late times. $n_{\textrm{s}}$ and $A_{\textrm{s}}$, on the other hand, relate to early universe cosmology. They are the  power-law spectral index and power of the primordial scalar perturbations respectively, evaluated at the pivot scale of $k_{*} = 0.05h$ Mpc$^{-1}$.  As defined in the previous section, the seventh parameter, $m_0$ is the mass of the lightest neutrino and it is the parameter of primary concern in this paper.

\item ii) The $\Lambda\textrm{CDM}+\sum m_{\nu}+r$ model: Apart from the main 7 parameters, in this model we also include the evolution of tensor perturbations in the analysis along with scalar perturbations, and add an additional free parameter $r$, which is the tensor-to-scalar ratio evaluated at the same pivot scale as $n_s$ and $A_s$.

\item iii) The $w\textrm{CDM}+\sum m_{\nu}$ model: Here instead of a cosmological constant with a dark energy equation of state (DE EoS hereafter) fixed at $w(z) = -1$ we opt for a DE EoS  $w$ which varies as a free parameter, but does not vary in time (i.e. $w$ can assume different values but the values are fixed throughout the evolution history of the universe). Here $z$ denotes the cosmological redshift ($z = 1/a -1$, where $a$ is the scale factor of FRW metric). Here, as well as in all models including non-$\Lambda$ dark energy, we use the PPF prescription \cite{Hu:2007pj} for incorporating dark energy perturbations.

\item iv) The $w_0 w_a \textrm{CDM}+\sum m_{\nu}$ model:  In this case we incorporate a dynamically varying DE EoS, i.e. $w(z)$ also varies with time. We parametrize the EoS with the $w_0-w_a$ approach. This is the Chevallier-Polarski-Linder (CPL) parametrization \cite{Chevallier:2000qy,Linder:2002et}:
\begin{equation} \label{eq5}
w(z) = w_0 + w_a (1-a) = w_0 + w_a \frac{z}{1+z}.
\end{equation}
We hereafter may simply refer to this model as DDE.  

\item v) The $w_0 w_a \textrm{CDM}+\sum m_{\nu}$ model with $w(z)\geq -1$: This model has the same parametrization as the previous model, but the dynamical dark energy is forced to stay in the non-phantom range, i.e. $w(z) \geq -1$. This is achieved by noticing that $w(z)$ in eq. (\ref{eq5}) is a monotonic function, that at present day $w(z) = w_0$ (since $a = 1$ is the present value of the scale factor by convention), and that at the very early universe ($a \rightarrow 0$) we had $w(z)\rightarrow w_0 + w_a$. It thus suffices to have the following hard priors applied on the CPL parameters to keep them from crossing the phantom barrier \cite{Choudhury:2018byy,Choudhury:2018adz, Vagnozzi:2018jhn}:
\begin{equation}
w_0 \geq -1;~~~~~~~~~~~~~~~~w_0+w_a \geq -1
\end{equation} 
We hereafter may simply refer to this model as NPDDE. 

\item vi) The $\Lambda \textrm{CDM} + \sum m_{\nu} + A_{\textrm{Lens}}$ model: In this extended model we include $A_{\textrm{Lens}}$, which is the scaling of the lensing amplitude. In a particular model, the theoretical prediction for the gravitational potential (which generates the weak lensing of the CMB) corresponds to $A_{\textrm{Lens}} = 1$. When $A_{\textrm{Lens}}$ is varied, the weak lensing is uncoupled from the primary anisotropies which cause it, and then scaled by the value of $A_{\textrm{Lens}}$ \cite{Calabrese:2008rt}. $A_{\textrm{Lens}}$ acts as a consistency check parameter. In a particular model, if the data prefers $A_{\textrm{Lens}} > 1$, it means it prefers more smoothing of its acoustic peaks in the power spectra (typically caused by lensing) than what theoretically should be.  The physical reason for this extra smoothing (if it can't be accounted for statistical fluctuation in the data) may not be extra lensing but may be any new effect that mimics lensing \cite{Aghanim:2018eyx}. There is a well-known $A_{\textrm{lens}}$ tension in the Planck high-$l$ data, but not in its measurements of lensing. In the $\Lambda\textrm{CDM}+A_{\rm Lens}$ model, with Planck 2018 TT+TE+EE+lowE data, one finds that the constraint  $A_{\textrm{Lens}}=1.180\pm0.065$ (68\%) \cite{Aghanim:2018eyx} is more than 2$\sigma$ away from $A_{\rm Lens} = 1$. Solving the $A_{\textrm{Lens}}$ anomaly with a statistical significance of 5$\sigma$ or more requires more precise measurement of the E and B mode polarization of the CMB photons in small scales, which will be possible with future CMB experiments like LiteBIRD \cite{Matsumura:2013aja} (space based); and stage III ground based experiments, provided the accurate measurement of $\tau$ from Planck is included. Optimistic Stage IV CMB experiments can even provide a resolution at 10$\sigma$, also providing significant information on scale dependance of $A_{\textrm{Lens}}$, if any. See \cite{Renzi:2017cbg} for details.

\item vii) The $\Lambda \textrm{CDM} + \sum m_{\nu} + \Omega_k$ model: Here we go away from a flat universe to the one which can be curved. The  curvature of the universe is parametrized by $\Omega_k$, which is called the curvature energy density, and we allow it to vary freely in this model. 

\end{itemize}

We use the publicly available Markov Chain Monte-Carlo package CosmoMC \cite{Lewis:2002ah} (which uses the Boltzmann solver CAMB \cite{Lewis:1999bs}) to perform a Bayesian analysis of cosmological datasets and derive constraints on $\sum m_{\nu}$ and other cosmological parameters. We use the Gelman and Rubin statistics \cite{doi:10.1080/10618600.1998.10474787} to estimate the convergence of the chains. All our chains had achieved the convergence criterion of $R-1 < 0.01$. We use flat priors on all the the parameters that are varied in a particular model. The priors are listed in table \ref{tab1}. 

We also calculate Bayesian evidences for the purpose of Bayesian model comparison between the NH and IH cases in the above cosmological models using the publicly available package CosmoChord \cite{Handley}, an extension to CosmoMC using PolyChord \cite{Handley:2015fda,10.1093/mnras/stv1911} for nested sampling. We used sufficiently broad priors on the cosmological parameters and 2000 live points for each CosmoChord run to keep the error to the estimated evidence small. The methodology to quantify evidence against IH closely follows \cite{Hannestad:2016fog}. The results are given in section \ref{sec5}. 

\begin{table}
\begin{center}
\begin{tabular}{c c}
\hline
Parameter                    & Prior\\
\hline
$\omega_c$         & [0.005, 0.1]\\
$\omega_b$         & [0.001, 0.99]\\
$\Theta_{\rm s}$             & [0.5, 10]\\
$\tau$                       & [0.01, 0.8]\\
$n_{\textrm{s}}$                        & [0.8, 1.2]\\
$\textrm{ln}[10^{10}A_{\textrm{s}}]$         & [2, 4]\\ 
$m_0$ (eV)                    & [0, 1]\\
$r$                          & [0, 1] \\
$w$                          & [-3, -0.33] \\
$w_0$                        & [-3, -0.33]\\
$w_a$                        & [-2, 2]\\ 
$\Omega_k$                   & [-0.3, 0.3] \\ 
$A_{\textrm{Lens}}$          & [0, 3]\\
\hline
\end{tabular}
\end{center}
\caption{\label{tab1} Flat priors on the main cosmological parameters constrained in this paper in the analyses with CosmoMC.}
\end{table}

\subsection{Datasets}\label{sec2.2}

~~~~~\textbf{CMB: Planck 2018.} We use the high-$l$ (30 $\leq$ $l$ $\leq$ 2508) and low-$l$ (2 $\leq$ $l$ $\leq$ 29) CMB TT likelihood along with high-$l$ E mode polarization and temperature-polarisation cross correlation likelihood from the recent Planck 2018 public release \cite{Aghanim:2019ame} and we call this combination ``TTTEEE". We use the Planck low-$l$ E mode polarization data, and in the text we mention it as ``lowE."  

\textbf{CMB: Planck 2018 lensing.} While the CMB anisotropy power spectra is determined from 2-point correlation functions (TT, TE, EE), the power spectra of the lensing potential is proportional to the 4-point correlation functions such as TTTT, TTEB and so on \cite{Aghanim:2018eyx}. We use it in our analyses as an additional CMB probe to ascertain neutrino physics properties, as lensing of CMB photons is produced by the gravitational potential of large scale structure which in turn is affected greatly by the free streaming massive neutrinos. We will refer to this dataset simply as ``lensing'' from now on. 

\textbf{CMB: BICEP2/Keck array data.} While running an MCMC analysis for the $\Lambda\textrm{CDM}+\sum m_{\nu}+r$ model, we also use the latest publicly available data (taken up to and including 2015) on the CMB BB spectra from the BICEP2/Keck collaboration (spanning the range: $20 < l < 330$) \cite{Ade:2018gkx}. This dataset is referred to as ``BK15" in the paper. 

\textbf{Baryon acoustic oscillations (BAO) Measurements.} In this paper we use the latest measurements of the BAO signal from different galaxy surveys: SDSS-III BOSS DR12 (galaxy samples at the effective redshifts of $z_{\textrm{eff}} =$ 0.38, 0.51 and 0.61) \cite{Alam:2016hwk}, the DR7 Main Galaxy Sample (MGS) at $z_{\textrm{eff}} = 0.15$ \cite{Ross:2014qpa}, and the Six-degree-Field Galaxy Survey (6dFGS) survey at $z_{\textrm{eff}} = 0.106$ \cite{Beutler:2011hx}). We simply name this combined dataset as ``BAO". 

All the CMB and BAO data together constitute our ``base'' dataset: \begin{center}{\textbf{Base} $\equiv$ \textbf{Planck 2018 TTTEEE + lowE + lensing + BAO.}} \end{center} 
For the $\Lambda\textrm{CDM}+\sum m_{\nu}+r$ model, the ``base" dataset shall also include BK15.

\textbf{Supernovae luminosity distance measurements.} We use the most recent Supernovae Type-Ia (SNe Ia) luminosity distance measurements from the Pantheon Sample \cite{Scolnic:2017caz}, which consists of distance information of 1048 SNe Ia ($0.01<z<2.3$), largest till date. Out of the 1048, 279 are from the Pan-STARRS1 (PS1) Medium Deep Survey  ($0.03 < z < 0.68$) and rest of them from SDSS, SNLS, various low-$z$ and HST samples. We call this dataset ``SNe".

To understand how the neutrino mass affects the cosmological observables measured by the above datasets and how we arrive at strong constraints on the same using these datasets, we refer the reader to \cite{Lattanzi:2017ubx}. 
\section{Results}\label{sec3}
In this section we provide the results of our analyses on the bounds on neutrino masses considering the three different hierarchies (degenerate, normal, and inverted). In section \ref{sec3.1} we discuss the results in the minimal $\Lambda\textrm{CDM}+\sum m_{\nu}$ model. The results in the extended models is discussed in section \ref{sec3.2}. Details about models and datasets are given in section \ref{sec2.1} and \ref{sec2.2} respectively. All the marginalized limits are given at 68\% C.L. (1$\sigma$), whereas cases where only upper or lower bounds are available, bounds are given at 95\% C.L. ($2\sigma$). The main results are contained in tables \ref{tab2}-\ref{tab5}.

\subsection{Results in the minimal $\Lambda\textrm{CDM} + \sum m_{\nu}$ model}
\label{sec3.1}

In this subsection we provide results of our analyses in the $\Lambda\textrm{CDM} + \sum m_{\nu}$ model, considering three different hierarchies (degenerate, normal, and inverted) for two dataset combinations, namely Base and Base+SNe, where Base $\equiv$ Planck 2018 TTTEEE+lowE+lensing+BAO. The main results are contained in table \ref{tab2} which provides confidence limits on cosmological parameters for the Base and Base+SNe combinations. 

\begin{table*}\centering
\ra{1.3}

\resizebox{\textwidth}{!}{\begin{tabular}{@{}rrrrcrrr@{}}\toprule
& \multicolumn{3}{c}{Base} & \phantom{abc} & \multicolumn{3}{c}{Base+SNe}\\
\cmidrule{2-4} \cmidrule{6-8}
& DH & NH & IH && DH & NH & IH\\ \midrule
$\Lambda\textrm{CDM}+\sum m_{\nu}$\\
$\omega_c$ & $0.1194\pm0.0009$ & $0.1192\pm 0.0009$ & $0.1191\pm 0.0009$ && $0.1193\pm 0.0009$ &  $0.1191\pm 0.0009$ & $0.1189\pm 0.0009$\\

$\omega_b$ & $0.02242\pm 0.00013$ & $0.02242^{+0.00013}_{-0.00014} $ & $0.02243\pm 0.00013$&& $0.02243\pm 0.00013$& $0.02244\pm 0.00013$ & $0.02244\pm 0.00013$ \\

$\Theta_{\textrm{s}}$ & $1.04100\pm 0.00029$ & $1.04100\pm 0.00029$ & $1.04100\pm 0.00029$&& $1.04102\pm 0.00029$ & $1.04103\pm 0.00029$ & $1.04103\pm 0.00029$\\

$\tau$ & $0.0554^{+0.0068}_{-0.0076}$ & $0.0569^{+0.0066}_{-0.0076}$& $0.0585^{+0.0069}_{-0.0076}$&& $0.0556\pm0.0071$ &$0.0573^{+0.0069}_{-0.0076}$ & $0.0588^{+0.0068}_{-0.0077}$\\

$n_{\textrm{s}}$ & $0.9666\pm 0.0036$ & $0.9668\pm 0.0037$&$0.9671\pm 0.0037$&&  $0.9669\pm 0.0036$& $0.9673\pm 0.0036$ & $0.9675\pm 0.0037$\\

$\textrm{ln}[10^{10}A_{\textrm{s}}]$ & $3.048^{+0.014}_{-0.015}$ & $3.051^{+0.014}_{-0.015}$ & $3.053 \pm 0.015$&& $3.046\pm0.014$ & $3.049\pm 0.014$ & $3.052^{+0.014}_{-0.015}$ \\

$m_0$ (eV) & $<0.040$ & $<0.040$& $<0.042$&&  $<0.038$ & $<0.038$& $<0.039$\\

$\sum m_{\nu}$ (eV) & $<0.12$& $<0.15$& $<0.17$ &&  $<0.11$ & $<0.14$ & $<0.16$\\

$H_0$ (km/s/Mpc) & $67.81^{+0.54}_{-0.46}$ & $67.50^{+0.49}_{-0.44}$& $67.22\pm0.45$&&$67.89^{+0.52}_{-0.45}$& $67.59\pm0.44$ & $67.33\pm 0.43$\\

$\sigma_8$ & $0.814^{+0.010}_{-0.007}$ & $0.806^{+0.009}_{-0.006}$& $0.799^{+0.008}_{-0.006}$&& $0.815^{+0.010}_{-0.007}$ & $0.806^{+0.008}_{-0.006}$ & $0.799^{+0.008}_{-0.006}$\\

$S_8$ & $0.827\pm0.011$ & $0.823\pm0.011$& $0.820\pm0.011$&&  $0.826\pm0.011$ & $0.822\pm0.011$ &  $0.818\pm0.011$\\
\midrule

$\Delta \chi^2 = \chi^2 - \chi^2_{IH}$ & $-2.89$ & $-0.95$ & 0&& $-2.73$ & $-1.27$ &  $0$\\
\bottomrule
\end{tabular}}
\caption{ Constraints on the cosmological parameters in the minimal $\Lambda\textrm{CDM}+\sum m_{\nu}$ model considering three different hierarchies (degenerate, normal, and inverted) with the Base and Base+SNe datasets, where Base $\equiv$ Planck 2018 TTTEEE+lowE+lensing+BAO.  Here $m_0$ is the mass of the lightest neutrino in a particular hierarchy and a freely varying parameter in the model, whereas $\sum m_{\nu}$, $H_0$, $\sigma_8$, and $S_8$ are derived parameters. Marginalized constraints are given at 1$\sigma$ whereas upper or lower bounds are given at 2$\sigma$. The $\chi^2$ differences are calculated at best-fit points. Details about models and datasets are given in section \ref{sec2}.}\label{tab2}
\end{table*}

\begin{figure}[tbp]
\centering 
\includegraphics[width=.4963\linewidth]{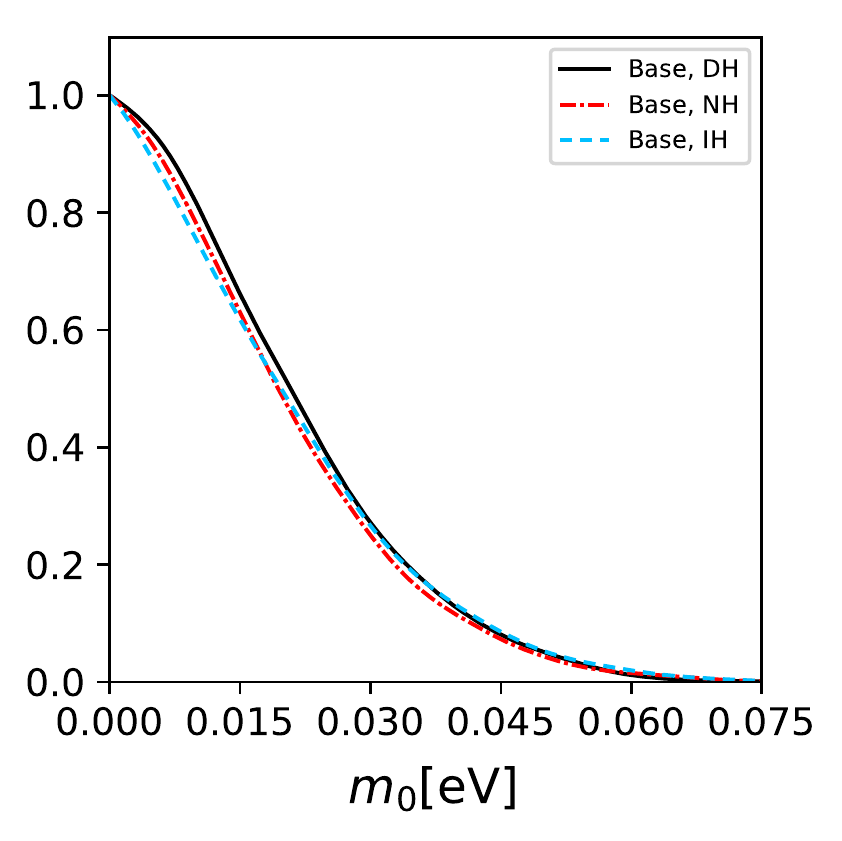}
\hfill
\includegraphics[width=.4963\linewidth]{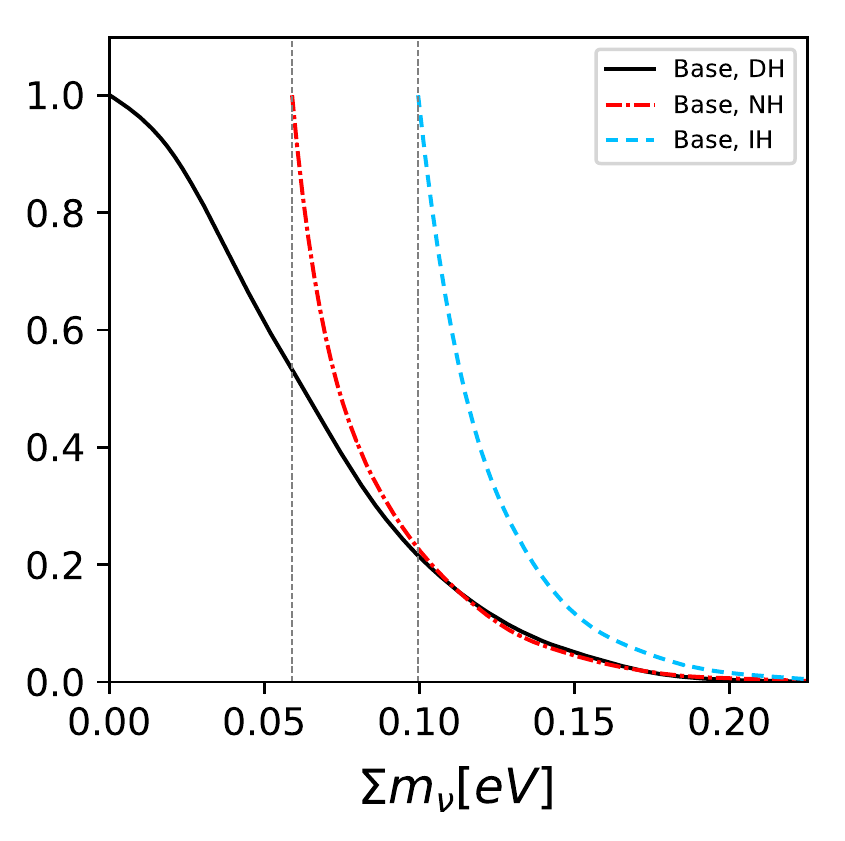}
\caption{\label{fig:1}Comparison of 1-D marginalized posterior distributions for $m_0$ (eV) and $\sum m_{\nu}$ (eV)  for the Base dataset in the $\Lambda\textrm{CDM}+\sum m_{\nu}$ model for the three hierarchies. The two dashed vertical lines are at $\sum m_{\nu} = 0.06$ and 0.10 eV are representing the minimum mass required for normal and inverted hierarchy respectively. }
\end{figure}

In figure \ref{fig:1} we depict the 1-D posterior distributions of $m_0$ (mass of the lightest neutrino in a hierarchy) and $\sum m_{\nu}$ in the $\Lambda\textrm{CDM}+\sum m_{\nu}$ model for Base and Base+SNe datasets considering different hierarchies. With the Base data recover the following 95\% bound on the mass sum: $\sum m_{\nu} < 0.12$ eV which is same as the bound of $\sum m_{\nu} < 0.12$ eV quoted by the Planck 2018 collaboration \cite{Ade:2018gkx} with the same data. The Planck 2015 result with similar data was $\sum m_{\nu} < 0.17$ eV (95\%, Planck 2015 TT,TE,EE+lowP+BAO). The main reason this improvement happens with Planck 2018 is because of improved measurement of $\tau$. Planck 2015 TT,TE,EE+lowP likelihoods produced a bound of $\tau = 0.079\pm 0.017$ (68\%) in the base $\Lambda\textrm{CDM}$ model \cite{Ade:2015xua}, which improved to $\tau = 0.0544^{+0.0070}_{-0.0081}$ (68\%) with Planck 2018 TT,TE,EE+lowE \cite{Aghanim:2018eyx}. The parameters $\tau$ and $\sum m_{\nu}$ are strongly correlated in the Planck TT data and high-$l$ polarization data \cite{Choudhury:2018byy}, and this degeneracy can be broken through better measurement of $\tau$ from the low-$l$ polarization data. Since an increase in $\sum m_{\nu}$ leads to suppression of the matter power spectrum which in turn leads to less gravitational lensing of the CMB photons, we see an increase in $\sum m_{\nu}$ leads to a decrease in the smearing of the CMB acoustic peaks (smearing happens due to lensing). This effect due to increasing neutrino masses, however, can be countered with increasing $\tau$, which exponentially suppresses power in the CMB anisotropy spectra (given other parameters are kept fixed). However, the effect of reionization is also found in the low-$l$ polarization data (TE, EE, BB) in the form of a ``reionization bump" whose amplitude is proportional to $\tau^2$ in the EE and BB spectra, and to $\tau$ in the TE spectra, and this change can't be compensated by varying other parameters in the model \cite{Reichardt2016}. Hence the improved removal of systematics in the low-$l$ polarization data with Planck 2018 helps in breaking the degeneracy between $\tau$ and $\sum m_{\nu}$, and this leads to stronger upper bounds on $\sum m_{\nu}$. However, not only $\tau$, Planck 2018 also largely corrects various systematic effects previously present in the high-$l$ polarization spectra of Planck 2015, and that also contributes towards obtaining a better bound on the $\sum m_{\nu}$. 

In our Base dataset combination, BAO is another effective tool in constraining neutrino masses. BAO data is useful in breaking the strong anti-correlation between the Hubble constant $H_0$ and $\sum m_{\nu}$ present in the Planck data. The comoving distance to the last scattering surface in a flat $\Lambda\textrm{CDM}+\sum m_{\nu}$ universe is defined as: $\chi(z_{dec}) = \int^{z_{dec}}_0 dz/H(z)$, where $z_{dec}$ is the redshift of photon decoupling, and $H(z) = \sqrt{\omega_{\gamma}(1+z)^4 + (\omega_c + \omega_b)(1+z)^3 + \omega_{\Lambda}+\rho_{\nu}(z)h^2/\rho_{cr,0}}$ (note: $\omega_i = \Omega_i h^2$, and $i\equiv \gamma, c, b, \Lambda$ with  $\gamma\equiv$ photons, $c\equiv$ CDM, $b\equiv$ baryons, $\Lambda\equiv$ cosmological constant; $\rho_{\nu}(z)$ is the neutrino energy density at a redshift $z$, $\rho_{cr,0} = 3H_0^2/8\pi G$ is the critical density today). In a flat universe $\Omega_\Lambda = 1 - \Omega_{\gamma} - (\Omega_c + \Omega_b) - \Omega_{\nu}$ and at late times $\rho_{\nu}(z)h^2/\rho_{cr,0} = \Omega_{\nu}h^2 \propto \sum m_{\nu}$. Now, given that early universe physics remains unchanged, $\chi(z_{dec})$ is very well constrained through $\Theta_s$ which is the most well constrained parameter by the Planck CMB anisotropies. On the other hand, $\Omega_{\gamma}$ and $(\omega_c + \omega_b)$ are also well constrained by the data. Thus any change in $\chi(z_{dec})$ due to an increase in $H_0$ (or $h = H_0/100$ km/sec/Mpc) has to be compensated by a decrease in $\sum m_{\nu}$ and vice versa, and hence there is a large anti-correlation between $H_0$ and $\sum m_{\nu}$. Thus in the $\Lambda\textrm{CDM} + \sum m_{\nu}$ model, lower values of $H_0$ correspond to higher values of $\sum m_{\nu}$ and higher values of $H_0$ correspond to lower $\sum m_{\nu}$. BAO data breaks this degeneracy partially by rejecting the low $H_0$ values preferred by Planck data, well studied in previous literature \cite{Hou:2012xq,Vagnozzi:2017ovm,Choudhury:2018byy}. For instance, in $\Lambda\textrm{CDM} + \sum m_{\nu}$, Planck 2015 TT,TE,EE+lowP prefers a value of $H_0 = 66.17^{+1.96}_{-0.81}$ km/sec/Mpc, whereas TT,TE,EE+lowP+BAO prefers $H_0=67.67^{+0.54}_{-0.51}$ km/sec/Mpc \cite{Choudhury:2018byy}.   

Apart from CMB power spectra and BAO, our Base dataset also contains Planck 2018 lensing likelihoods, which as per Planck 2018 collaboration \cite{Aghanim:2018eyx} prefers a slightly increased lensing power spectrum amplitude compared to Planck 2015, and thus leads to a slightly tighter neutrino mass constraints, contrary to Planck 2015 lensing likelihoods which used to relax the constraints. In our case, without the Planck 2018 lensing likelihoods, we obtained a bound of $\sum m_{\nu} < 0.13$ eV (95\%, Planck 2018 TT,TE,EE+lowE+BAO), i.e., Planck 2018 lensing has only a small effect in this $\Lambda\textrm{CDM} + \sum m_{\nu}$ model when used along with Planck 2018 CMB anisotropies, and BAO.

\begin{figure}[tbp]
\centering 
\includegraphics[width=.3\linewidth]{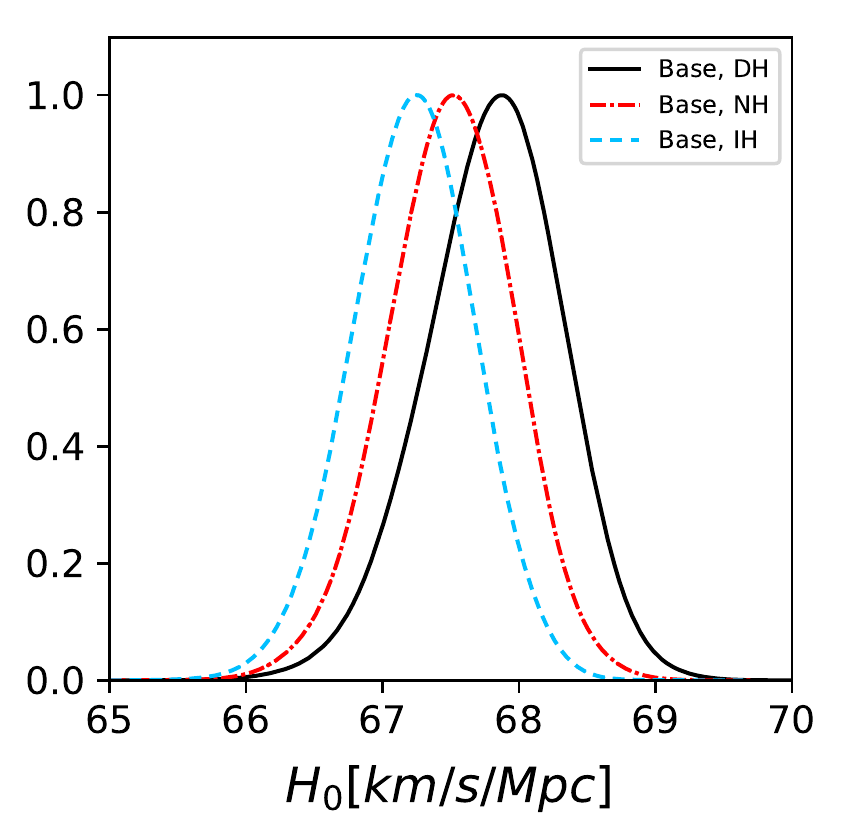}
\hfill
\includegraphics[width=.3\linewidth]{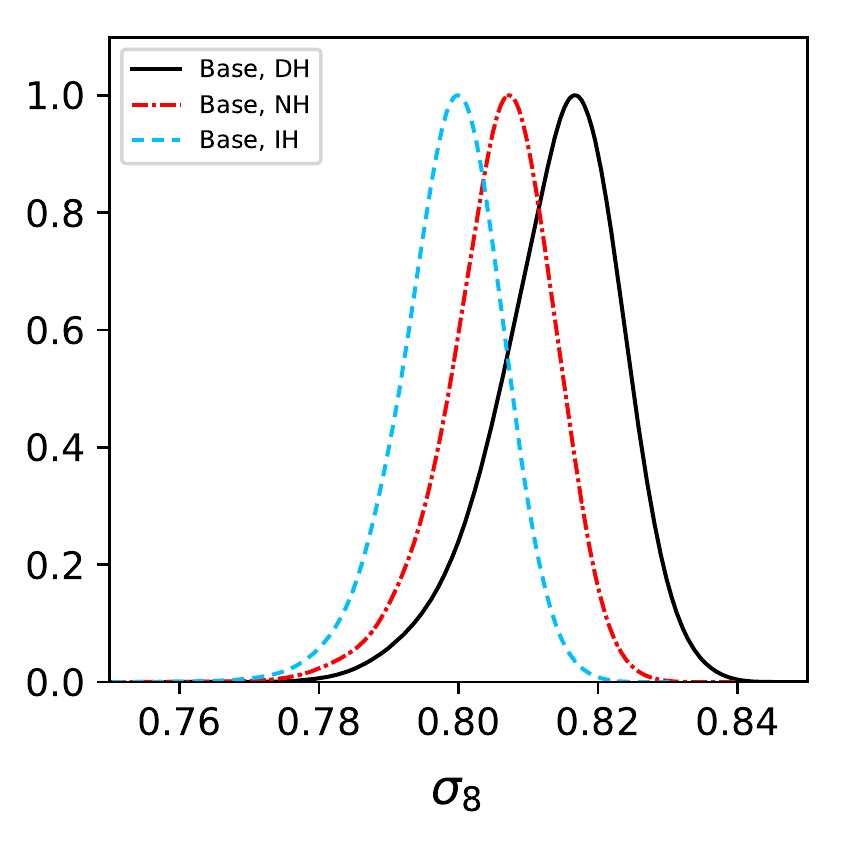}
\hfill
\includegraphics[width=.3\linewidth]{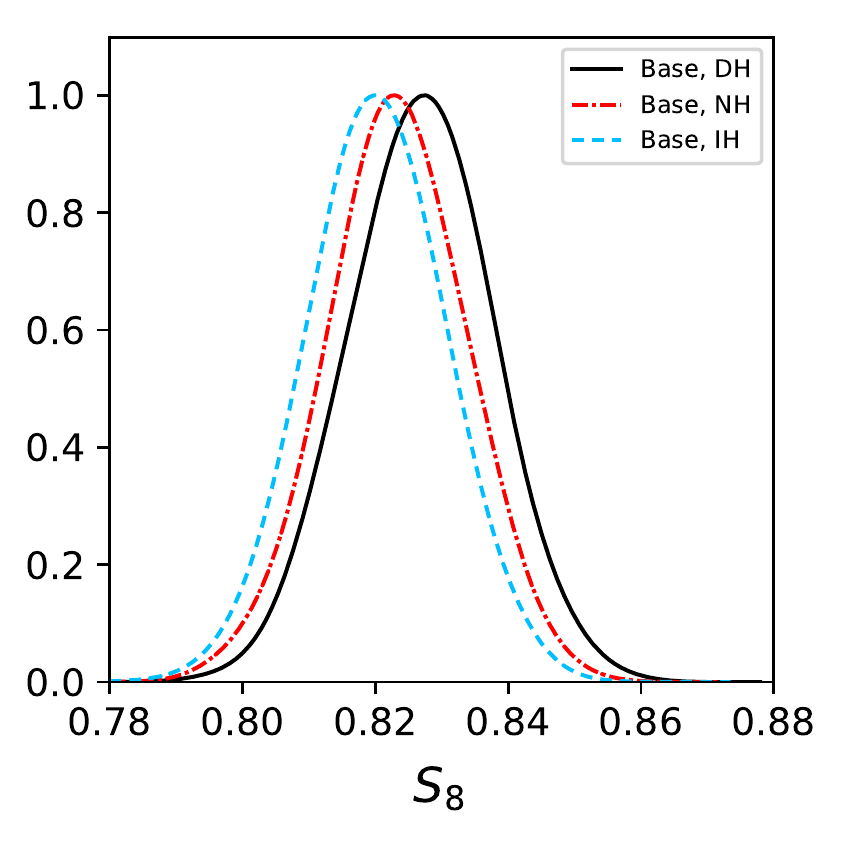}
\caption{\label{fig:2}Comparison of 1-D marginalized posterior distributions for $H_0$ (km/sec/Mpc), $\sigma_8$, and $S_8$ for the Base dataset in the $\Lambda\textrm{CDM}+\sum m_{\nu}$ model for the three hierarchies.}
\end{figure}

We see that the bounds on $\sum m_{\nu}$ do differ significantly across the different hierarchies. We see that with the Base dataset we get the following bounds: $\sum m_{\nu} < 0.15$ eV (NH), $\sum m_{\nu} < 0.17$ eV (IH) at 95\% C.L. As previously stated, we use the mass of the lightest neutrino, $m_0$ as the varying parameter and then use the mass-squared splittings given in eq. \ref{eq1} to determine the other masses in a particular hierarchy, and this implicitly puts a prior on the $\sum m_{\nu}$, i.e., $\sum m_{\nu} \geq 0.06$ eV for normal hierarchy, and $\sum m_{\nu} \geq 0.10$ eV for inverted hierarchy. The reason the bound on $\sum m_{\nu}$ differs significantly across the three hierarchies is possibly the priors imposed on $\sum m_{\nu}$ (see figure \ref{fig:1} for the visualization of the same). We cross-check this by running MCMC chains with the degenerate hierarchy, but with the priors: i)  $\sum m_{\nu} \geq 0.06$ eV, and ii) $\sum m_{\nu} \geq 0.10$ eV. We found a 95\% upper bound of $\sum m_{\nu} < 0.15$ eV in the first case, and in the second case it was $\sum m_{\nu} < 0.18$ eV. It seems evident that the priors do play an important role in relaxing the bounds. However it is to be noted that the method we have used in this paper for the normal and inverted hierarchy (i.e. with lightest mass $m_0$ and the mass-squared splittings from oscillations experiments) produces bounds which are slightly stronger than the degenerate case with the priors $\sum m_{\nu} \geq 0.06$ eV and $\sum m_{\nu} \geq 0.10$ eV respectively, i.e. the two methods are not completely equivalent. The difference is clearer with Planck 2018 TT,TE,EE+lowE data, with which the NH case leads to a 95\% bound of $\sum m_{\nu} < 0.29$ eV, whereas the DH case with the $\sum m_{\nu} \geq 0.06$ eV prior gives us a bound of $\sum m_{\nu} < 0.32$ eV. For the same data, the IH case produces a 95\% bound of $\sum m_{\nu} < 0.33$ eV,  whereas the DH case with the $\sum m_{\nu} \geq 0.10$ eV prior obtains a bound of $\sum m_{\nu} < 0.35$ eV. As discussed in the introduction, in case of NH and IH, with the lightest mass ($m_0$) parametrization and a uniform prior on $m_0$, the implicit prior on $\sum m_{\nu}$ is not completely flat or uniform (i.e. rises at the low values of $m_0$). This non-flat prior causes the low values of $\sum m_{\nu}$ to be more favoured compared to the case of a flat prior on $\sum m_{\nu}$, and hence, this can explain the slightly tighter bounds on $\sum m_{\nu}$ with the $m_0$ parametrization. 

From table \ref{tab2} we can see an important trend: in the $\Lambda\textrm{CDM}+\sum m_{\nu}$ model with Base dataset, as we go from degenerate to normal to inverted hierarchy there appears a decrease in the preferred values of $H_0$ and $\sigma_8$. This is also visualized in figure \ref{fig:2}. The reason for the decrease in $H_0$ is simply the anti-correlation between $H_0$ and $\sum m_{\nu}$ as described above, and the fact that the normal hierarchy prefers neutrino masses which are larger than the degenerate case, and the inverted hierarchy prefers masses which are larger than the normal hierarchy. There is, again, a strong anti-correlation present between  $\sigma_8$ and $\sum m_{\nu}$, since $\sigma_8$ is the amplitude of the linear matter power spectrum at a length scale of $8h^{-1}$MPc, and increasing $\sum m_{\nu}$ increases the neutrino energy density and that increases the suppression of the small scale matter power-spectrum leading to a lower $\sigma_8$. Thus, from degenerate to normal to inverted hierarchy, the $H_0$-tension between Planck and local measurements increases, whereas the $\sigma_8$-tension \cite{Kitching:2016hvn} between Planck and cosmic shear measurements decreases. This trend remains true for the other datasets we have considered in this model (as can be seen from table \ref{tab2}). 
From figure \ref{fig:2} we also observe that the parameter $S_8 = \sigma_8 \sqrt{\Omega_m/0.3}$ also goes downward, but not as strongly as $\sigma_8$, since increasing neutrino masses also cause $\Omega_m$ to increase which compensates for the decreased $\sigma_8$, i.e. $S_8$ is defined such that its correlation with $\sum m_{\nu}$ is small in this model.  Again, while $H_0$ and $\sigma_8$ both are strongly anti-correlated with $\sum m_{\nu}$, the magnitude of the correlation coefficients ($R_{ij} = C_{ij}/\sqrt{C_{ii}C_{jj}}$ where $i$,$j$ are the two parameters, and $C$ is the covariance matrix of the cosmological parameters) decrease from degenerate to normal to inverted hierarchy because of the implicit priors on $\sum m_{\nu}$, i.e the priors help in partially breaking the degeneracies. We find that $R_{H_0,\Sigma m_{\nu}} =-$0.55 (DH), $-$0.47 (NH), $-$0.43 (IH); and $R_{\sigma_8,\Sigma m_{\nu}} =-$0.76 (DH), $-$0.72 (NH), $-$0.67 (IH).

Apart from the results with the Base dataset, in table \ref{tab2}, we provide results with the Base+SNe dataset also. In the absence of a varying dark energy equation of state, the SNe data is able to constrain $\Omega_m$ effectively \cite{Scolnic:2017caz}, while the Planck CMB data can constrain $\Omega_m h^2$ well. Thus together they can effectively constrain $H_0$, and as found out in \cite{Choudhury:2018byy}, the Planck+SNe combination actually prefers $H_0$ values which are higher than Planck alone, and thus the SNe data can help in breaking the degeneracy between $H_0$ and $\sum m_{\nu}$ partially. BAO data is however much more efficient in breaking the degeneracy and with Base+SNe (note that Base already contains BAO). In the DH case, we find the following 95\% bounds on the neutrino mass sum in this $\Lambda\textrm{CDM}+\sum m_{\nu}$ model: $\sum m_{\nu}<0.11$ eV, which is only slightly stronger than the bound without the SNe data. 

\textbf{Akaike information criterion (AIC).} To compare the goodness of fit of different models to the same data  a popular method is to compute the  Akaike information criterion (AIC) \cite{1100705} for each model, defined as :
\begin{equation}
\textrm{AIC} = \chi^2_{\textrm{best-fit}} + 2k
\end{equation}
where $k$ is the number of parameters in the model. Comparison with another model is done by computing the difference: $\Delta \textrm{AIC} = \Delta \chi^2_{\textrm{best-fit}} + 2\Delta k$. Between two models the model with a lower AIC is considered the preferred model.  The $2\Delta k$ term penalizes the model with greater number of parameters as it is usually able to provide better fit because of the larger parameter space being available to it. In this work however, we are interested in the comparison of the quality of fits between the neutrino mass hierarchies and since we have the same number of parameters in all the different cases, $2\Delta k = 0$, and  $\Delta \textrm{AIC} = \Delta \chi^2_{\textrm{best-fit}}$. We have provided the $\Delta \textrm{AIC} = \chi^2_{\textrm{best-fit}} - \chi^2_{\textrm{IH,best-fit}}$ values for each neutrino hierarchy in table \ref{tab2}. We find that cosmological datasets slightly prefer the normal hierarchy over the inverted one. In this $\Lambda\textrm{CDM}+\sum m_{\nu}$ model, with the Base data, we find that the NH is preferred to the IH by a $\Delta \textrm{AIC} = -0.95$, which is a very mild result. 

This showcases that the current cosmological data is not sensitive enough to differentiate between the two hierarchies,
and is completely consistent with the findings in e.g. \cite{Hannestad:2016fog,Mahony:2019fyb} which both find that a formal sensitivity to $\sum m_\nu$ in the 0.01-0.02 eV range is required to guarantee a distinction between the two hierarchies in case the true value of $\sum m_\nu$ is sufficiently less than 0.1 eV (i.e. if the neutrino masses follow NH). This is not possible using current data, but will become possible in the near future using high precision data from e.g.\ EUCLID (see e.g. \cite{Hamann:2012fe,Brinckmann:2018owf,Sprenger:2018tdb,Chudaykin:2019ock} for discussions on this topic). However, if the true value of $\sum m_\nu > 0.1$ eV, we won't be able to tell one hierarchy from the other even with the said sensitivity.
With the other dataset combinations also the $\Delta \textrm{AIC}$ remains mild between NH and IH. The (unphysical) degenerate hierarchy on the other hand, produces a better fit to the data compared to both NH and IH in case of all the dataset combinations we have studied.

\subsection{Results in the extended models}
\label{sec3.2}

In this section we present the results in the extended cosmologies. Table \ref{tab4} contains the constraints on selected cosmological parameters in the extended models with Base and Base+SNe datasets. Marginalized constraints are given at 1$\sigma$ whereas upper or lower bounds are given at 2$\sigma$. Details about models and datasets are given in section \ref{sec2}. Note that the Base data also contains BK15 when we consider the $\Lambda\textrm{CDM}+\sum m_{\nu}+r$ model. 

\begin{table*}\centering
\ra{1.3} 
\resizebox{\textwidth}{!}{\begin{tabular}{@{}rrrrcrrr@{}}\toprule
& \multicolumn{3}{c}{Base} & \phantom{abc} & \multicolumn{3}{c}{Base+SNe}\\
\cmidrule{2-4} \cmidrule{6-8} 
& DH & NH & IH && DH & NH & IH \\ \midrule
$\Lambda\textrm{CDM}+\sum m_{\nu}+r$\\
$r$ & $<0.0626$ & $<0.0632$& $<0.0618$&& $<0.0610$& $<0.0629$& $<0.0624$\\
$m_0$ (eV) & $<0.038$ & $<0.039$& $<0.040$&& $<0.037$& $<0.035$& $<0.038$\\

$\sum m_{\nu}$ (eV) &$<0.11$& $<0.14$ & $<0.17$&& $<0.11$& $<0.13$& $<0.16$\\

$H_0$ (km/s/Mpc) & $67.78^{+0.52}_{-0.46}$& $67.48^{+0.48}_{-0.45}$& $67.21\pm0.44$&& $67.86\pm0.48$& $67.58^{+0.44}_{-0.45}$ & $67.30^{+0.43}_{-0.44}$\\

$\sigma_8$ & $0.815^{+0.010}_{-0.007}$& $0.807^{+0.008}_{-0.006}$& $0.800^{+0.008}_{-0.006}$&& $0.816^{+0.010}_{-0.007}$& $0.808^{+0.008}_{-0.007}$& $0.800^{+0.007}_{-0.006}$ \\

$S_8$ & $0.832\pm0.011$& $0.825^{+0.011}_{-0.010}$& $0.822\pm0.011$&& $0.828\pm0.011$& $0.824\pm0.010$& $0.820\pm0.010$ \\

\midrule

$w\textrm{CDM}+\sum m_{\nu}$\\
$w$ & $ -1.042^{+0.072}_{-0.052}$ & $-1.068^{+0.071}_{-0.052}$ & $-1.089^{+0.070}_{-0.055}$ && $ -1.025^{+0.037}_{-0.033}$& $ -1.037^{+0.036}_{-0.032}$ & $ -1.048\pm0.034$\\
$m_0$ (eV) & $<0.062$ & $<0.066$ & $<0.066$&& $<0.053$& $<0.053$& $<0.054$\\

$\sum m_{\nu}$ (eV) &  $<0.19$& $<0.21$ & $<0.23$&& $<0.16$& $<0.18$& $<0.20$\\

$H_0$ (km/s/Mpc) &  $68.67^{+1.33}_{-1.59}$& $69.01^{+1.31}_{-1.60}$& $69.24^{+1.38}_{-1.61}$&& $68.33\pm0.82$& $68.31\pm0.82$ & $68.27\pm0.82$\\

$\sigma_8$ & $0.821^{+0.016}_{-0.017}$& $0.820\pm0.016$& $0.819\pm0.016$&& $0.818^{+0.013}_{-0.011}$& $0.814\pm0.012$& $0.810\pm0.012$ \\

$S_8$ & $0.825\pm0.011$& $0.822^{+0.012}_{-0.011}$& $0.822\pm0.011$&& $0.826\pm0.011$& $0.823\pm0.011$& $0.821\pm0.011$ \\

\midrule

$w_0 w_a \textrm{CDM}+\sum m_{\nu}$ \\
$w_0$ & $-0.68^{+0.26}_{-0.14}$ &$-0.68^{+0.26}_{-0.14}$ & $-0.68^{+0.25}_{-0.13}$&&  $-0.94^{+0.08}_{-0.09}$& $-0.94\pm0.09$& $-0.93\pm0.09$\\
$w_a$ & $ -1.06^{+0.37}_{-0.79}$ & $<-0.085$& $<-0.164$   && $-0.41^{+0.46}_{-0.29}$& $-0.49^{+0.44}_{-0.30}$& $-0.56^{+0.43}_{-0.32}$\\
$m_0$ (eV) & $<0.083$ & $<0.080$& $<0.083$   && $<0.089$ & $<0.088$& $<0.088$\\

$\sum m_{\nu}$ (eV) &  $<0.25$& $<0.26$ & $<0.28$  && $<0.27$& $<0.28$& $<0.29$\\

$H_0$ (km/s/Mpc) &  $65.70^{+1.60}_{-2.47}$& $65.78^{+1.61}_{-2.47}$& $65.80^{+1.62}_{-2.43}$&& $68.28\pm0.83$& $68.23\pm0.84$ & $68.23\pm0.82$\\

$\sigma_8$ & $0.795^{+0.018}_{-0.023}$& $0.792^{+0.017}_{-0.023}$& $0.790^{+0.018}_{-0.023}$&& $0.817^{+0.015}_{-0.013}$& $0.813^{+0.014}_{-0.012}$& $0.811^{+0.013}_{-0.012}$ \\

$S_8$ & $0.837\pm0.014$& $0.834^{+0.014}_{-0.013}$ & $0.832\pm0.013$&& $0.827\pm0.012$& $0.826\pm0.012$& $0.824\pm0.012$ \\

\midrule

$w_0 w_a \textrm{CDM}+\sum m_{\nu}$($w(z)\geq -1)$\\
$w_0$ & $<-0.873$ & $<-0.888$& $<-0.900$&& $<-0.937$& $<-0.944$& $<-0.949$\\

$w_a$ & $0.009^{+0.057}_{-0.067}$ &$0.007^{+0.049}_{-0.058}$&$0.007^{+0.044}_{-0.050}$&& $0.028^{+0.034}_{-0.056}$& $0.022^{+0.029}_{-0.047}$ & $0.020^{+0.025}_{-0.043}$\\

$m_0$ (eV) & $<0.032$ & $<0.034$& $<0.035$&& $<0.032$& $<0.033$& $<0.035$\\

$\sum m_{\nu}$ (eV) &  $<0.10$& $<0.13$ & $<0.16$&& $<0.09$& $<0.13$& $<0.16$\\

$H_0$ (km/s/Mpc) & $66.64^{+0.97}_{-0.66}$& $66.46^{+0.88}_{-0.62}$& $66.33^{+0.83}_{-0.57}$&& $67.23^{+0.63}_{-0.53}$& $67.01^{+0.57}_{-0.51}$ & $66.81^{+0.54}_{-0.48}$\\

$\sigma_8$ & $0.801^{+0.012}_{-0.010}$& $0.795^{+0.011}_{-0.009}$& $0.789^{+0.010}_{-0.008}$&& $0.807^{+0.010}_{-0.008}$& $0.799^{+0.009}_{-0.008}$& $0.793\pm0.008$ \\

$S_8$ & $0.826\pm0.011$& $0.823\pm0.011$& $0.820\pm0.011$&& $0.824\pm0.010$& $0.821\pm0.011$& $0.817\pm0.011$ \\

\midrule
$\Lambda\textrm{CDM}+\sum m_{\nu}$+$A_{\textrm{Lens}}$\\

$A_{\textrm{Lens}}$ & $1.100^{+0.046}_{-0.056}$ & $1.107^{+0.042}_{-0.055}$& $1.116^{+0.040}_{-0.050}$&& $1.098^{+0.044}_{-0.058}$&$1.106^{+0.042}_{-0.053}$&$1.116^{+0.040}_{-0.050}$ \\

$m_0$ (eV) & $<0.098$ & $<0.094$& $<0.092$&& $<0.094$& $<0.090$& $<0.087$\\

$\sum m_{\nu}$ (eV) &  $<0.29$& $<0.29$ & $<0.30$&& $<0.28$& $<0.28$& $<0.29$\\

$H_0$ (km/s/Mpc) &  $67.76^{+0.71}_{-0.60}$& $67.66^{+0.66}_{-0.59}$& $67.56^{+0.63}_{-0.53}$&& $67.86\pm^{+0.68}_{-0.57}$& $67.76\pm0.59$ & $67.66^{+0.58}_{-0.51}$\\

$\sigma_8$ & $0.782^{+0.029}_{-0.018}$& $0.777^{+0.025}_{-0.013}$& $0.772^{+0.022}_{-0.012}$&& $0.784^{+0.028}_{-0.017}$&$0.779^{+0.024}_{-0.013}$& $0.773^{+0.021}_{-0.011}$ \\

$S_8$ & $0.793^{+0.023}_{-0.020}$& $0.790^{+0.022}_{-0.017}$& $0.786^{+0.020}_{-0.016}$&& $0.793^{+0.023}_{-0.019}$&$0.790^{+0.021}_{-0.017}$& $0.785^{+0.019}_{-0.016}$ \\
\midrule
$\Lambda\textrm{CDM}+\sum m_{\nu}+\Omega_{\textrm{k}}$\\

$\Omega_{\textrm{k}}$ &$0.0004\pm0.0020$ & $0.0012^{+0.0020}_{-0.0021}$& $0.0019^{+0.0019}_{-0.0021}$&& $0.0004\pm0.0021$& $0.0012\pm0.0020$& $0.0019\pm0.0020$\\

$m_0$ (eV) & $<0.050$  & $<0.051$& $<0.053$&& $<0.044$& $<0.047$& $<0.049$\\

$\sum m_{\nu}$ (eV) &  $<0.15$& $<0.17$ & $<0.20$&& $<0.13$& $<0.16$& $<0.19$\\

$H_0$ (km/s/Mpc) &  $67.87\pm0.67$& $67.75\pm0.67$& $67.67^{+0.67}_{-0.68}$&& $67.95\pm0.67$& $67.84\pm0.66$ & $67.77\pm0.66$\\

$\sigma_8$ & $0.813^{+0.011}_{-0.008}$& $0.807^{+0.010}_{-0.008}$& $0.801^{+0.009}_{-0.008}$&& $0.814^{+0.010}_{-0.009}$& $0.807^{+0.010}_{-0.008}$& $0.801^{+0.009}_{-0.008}$ \\

$S_8$ & $0.826\pm0.011$& $0.823\pm0.011$& $0.819\pm0.011$ && $0.825\pm0.011$& $0.822\pm0.011$& $0.818\pm0.010$ \\

\bottomrule
\end{tabular}}
\caption{Constraints on selected cosmological parameters in the extended models considering three different hierarchies (degenerate, normal, and inverted) with the Base and Base+SNe datasets. In the $\Lambda\textrm{CDM}+\sum m_{\nu}+ r$ model Base data also includes BK15.}\label{tab4}
\end{table*}
\begin{figure}[tbp]
\centering 
\includegraphics[width=.4963\linewidth]{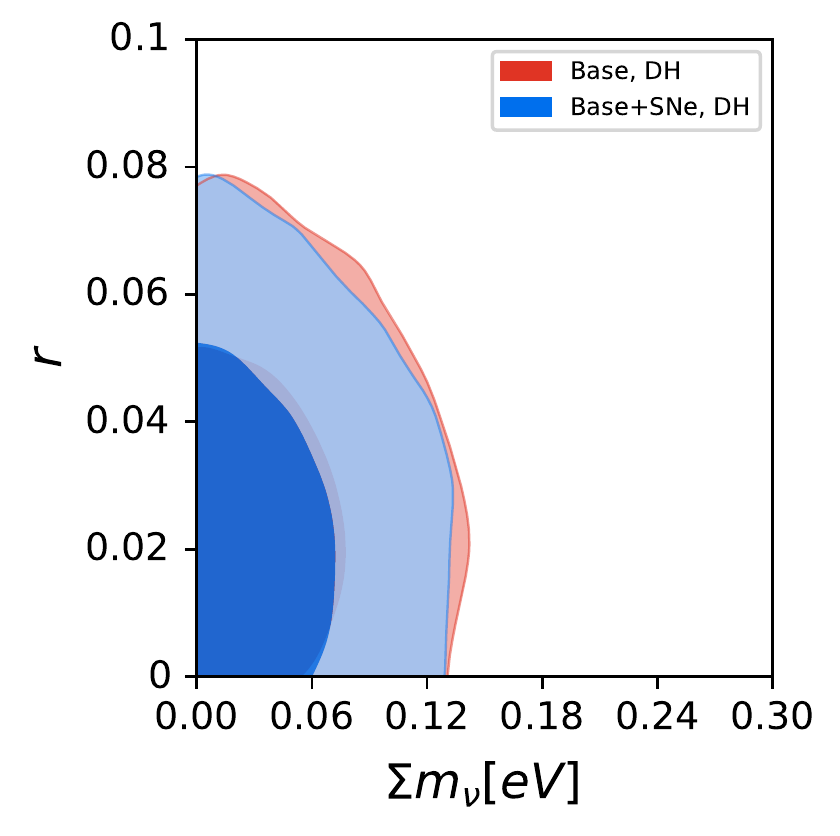}
\hfill
\includegraphics[width=.4963\linewidth]{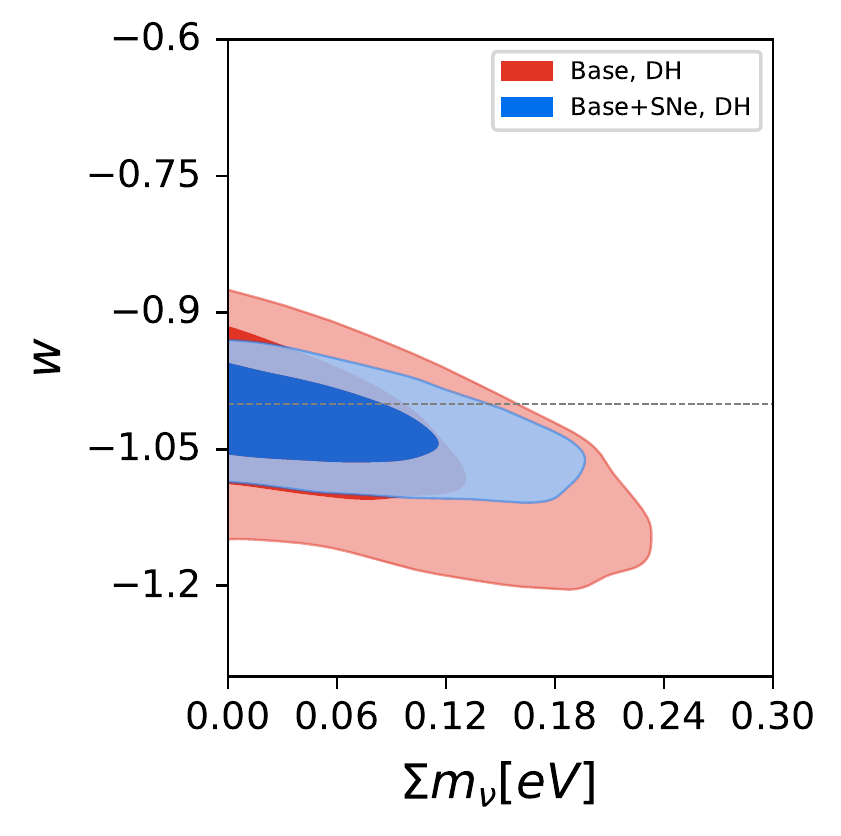}
\caption{\label{fig:3} 68\% and 95\% marginalized contours for $r$ vs $\sum m_{\nu}$ in the $\Lambda\textrm{CDM}+\sum m_{\nu}+ r$ model on the left, and for $w$ vs $\sum m_{\nu}$ in the $w\textrm{CDM}+\sum m_{\nu}$ model on the right, using Base and Base+SNe datasets and considering degenerate hierarchy. For the $\Lambda\textrm{CDM}+\sum m_{\nu}+ r$ model Base data also includes BK15 data. }
\end{figure}

\begin{itemize}
\item i) \textbf{Results in the $\Lambda\textrm{CDM}+\sum m_{\nu}+r$ model:} In this model we allow the tensor perturbations to vary as well, along with scalar ones. The CMB B-mode polarization comes from two sources: i) primordial gravitational waves, ii) gravitational weak lensing. With the Base data (which also includes BK15) we get the following 95\% bounds: $\sum m_{\nu} < 0.11$ eV (DH), $\sum m_{\nu} < 0.14$ eV (NH), and $\sum m_{\nu} < 0.17$ eV (IH). We do not see any big changes on the mass bounds compared to $\Lambda\textrm{CDM}+\sum m_{\nu}$. The bounds are, however, slightly tighter in the $\Lambda\textrm{CDM}+\sum m_{\nu}+r$ model with the BK15 data than without, most likely due to the additional lensing information encoded in the B mode data BK15, since $r$ and $\sum m_{\nu}$ are not correlated, as can be seen from the left panel of figure \ref{fig:3}. 
The effect of the neutrino mass bounds getting slightly tighter with B mode data was previously noticed in \cite{Choudhury:2018byy,Choudhury:2018adz,Choudhury:2018sbz} with BK14 data \cite{Array:2016afx}, predecessor of BK15. 

\item ii) \textbf{Results in the $w\textrm{CDM}+\sum m_{\nu}$ model:} In this model we go away from the cosmological constant, and consider the DE EoS $w$ as a varying parameter. A well-known degeneracy exists between $w$ and $\sum m_{\nu}$ \cite{Hannestad:2005gj}, through their mutual degeneracy with $\Omega_m$, which considerably relaxes the bounds on $\sum m_{\nu}$ in this model compared to $\Lambda\textrm{CDM}+\sum m_{\nu}$. The visualization of the anti-correlation between the two parameters is available in the right panel of figure \ref{fig:3} for the Base and Base+SNe data. We find a correlation coefficient of $R_{w,\sum m_{\nu}} = -$0.57 (DH), $-$0.52 (NH), $-$0.47 (IH) with the Base data, whereas $R_{w,\sum m_{\nu}} = -$0.45 (DH), $-$0.35 (NH), $-$0.30 (IH) with Base+SNe. The SNe data when used with the Base data, reduces the magnitude of the anti-correlation with better constraints on $w$. This happens, since in the SNe data $w$ and $\Omega_m$ are strongly anti-correlated, whereas in the CMB data $w$ and $\Omega_m$ are strongly correlated (for instance see figure 20 of \cite{Scolnic:2017caz}). So including the SNe data with Base leads to a large decrease in the $w$-$\Omega_m$ correlation, which in turn leads to a decrease in magnitude of $R_{w,\sum m_{\nu}}$. Also from figure \ref{fig:3}, we see that due to the anti-correlation lower values of $w$ prefer higher $\sum m_{\nu}$ and higher values of $w$ prefer lower $\sum m_{\nu}$. The Base+SNe combination constraints $w$ better than Base such that the lowest values of $w$ allowed by Base data is rejected, and this leads to stronger bounds on $\sum m_{\nu}$ with Base+SNe, as can be seen from table \ref{tab4}. The SNe data also rejects high $w$ values, but those regions prefer low $\sum m_{\nu}$ values and hence rejecting the high $w$ region does not help in strengthening the upper bound on $\sum m_{\nu}$. 

There is also a strong degeneracy between $w$ and $H_0$, that exists since lower values of $w$ correspond to higher present day expansion rate and hence higher $H_0$ values. This happens since changing $w$ can change the comoving distance to the last scattering surface $\chi(z_{dec})= \int^{z_{dec}}_0 dz/H(z)$, which is well-constrained by the CMB data, and thus any change in $\chi(z_{dec})$ needs to be compensated with another parameter. We have,  
\begin{equation}
H(z) = \sqrt{\omega_{\gamma}(1+z)^4 + (\omega_c + \omega_b)(1+z)^3 + \Omega_{DE}(z)h^2+\rho_{\nu}(z)h^2/\rho_{cr,0}}.
\end{equation}

Here $\Omega_{DE}(z) = \Omega_{DE}(0) (1+z)^{3(1+w)}$ is the energy density of dark energy with EoS $w$, and at late times when dark energy is a dominant component, decreasing the value of $w$ leads to a decrease in $H(z)$, which can be compensated by increasing either $h$ (or $H_0$) or $\sum m_{\nu}$ or both. This is why in the $w\textrm{CDM}+\sum m_{\nu}$ model, with Base data, we find that not only $w$ and $H_0$ are anti-correlated ($R_{w,H_0} =-0.91$) (DH), the correlation between $H_0$ and $\sum m_{\nu}$ is inverted from what it was in $\Lambda\textrm{CDM}+\sum m_{\nu}$, i.e. here $H_0$ and $\sum m_{\nu}$ are positively correlated with  $R_{H_0,\sum m_{\nu}} =+0.30$ (DH). The SNe data, when combined with Base, constrains $w$ and $H_0$ in a much better way and breaks the degeneracy present between $H_0$ and $\sum m_{\nu}$, and we find that $R_{H_0,\sum m_{\nu}} =-0.013$ (DH) with Base+SNe. This is why, in table \ref{tab4} with the Base data, as we go from DH to NH to IH, the preference for higher and higher neutrino masses and the positive correlation between $H_0$ and $\sum m_{\nu}$ leads to slight increase in the preferred values of $H_0$, but with Base+SNe, the almost negligible correlation explains the lack of any big change in the preferred $H_0$ values across the different hierarchies. On the other hand, the strong degeneracy between $w$ and $H_0$ survives, but becomes slightly weaker at $R_{w,H_0} =-0.76$ (DH) with Base+SNe.  This leads to the $w\textrm{CDM}+\sum m_{\nu}$ model predominantly preferring values of $w<-1$, i.e. the phantom DE region. 
\end{itemize}

\begin{figure}[tbp]
\centering 
\includegraphics[width=.3\linewidth]{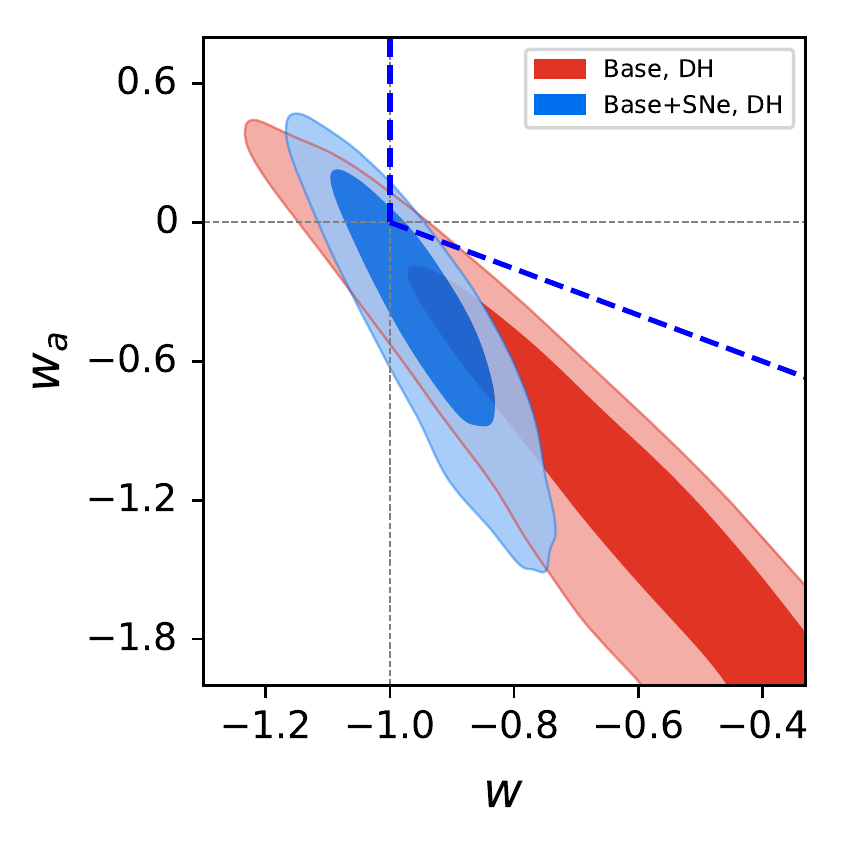}
\hfill
\includegraphics[width=.3\linewidth]{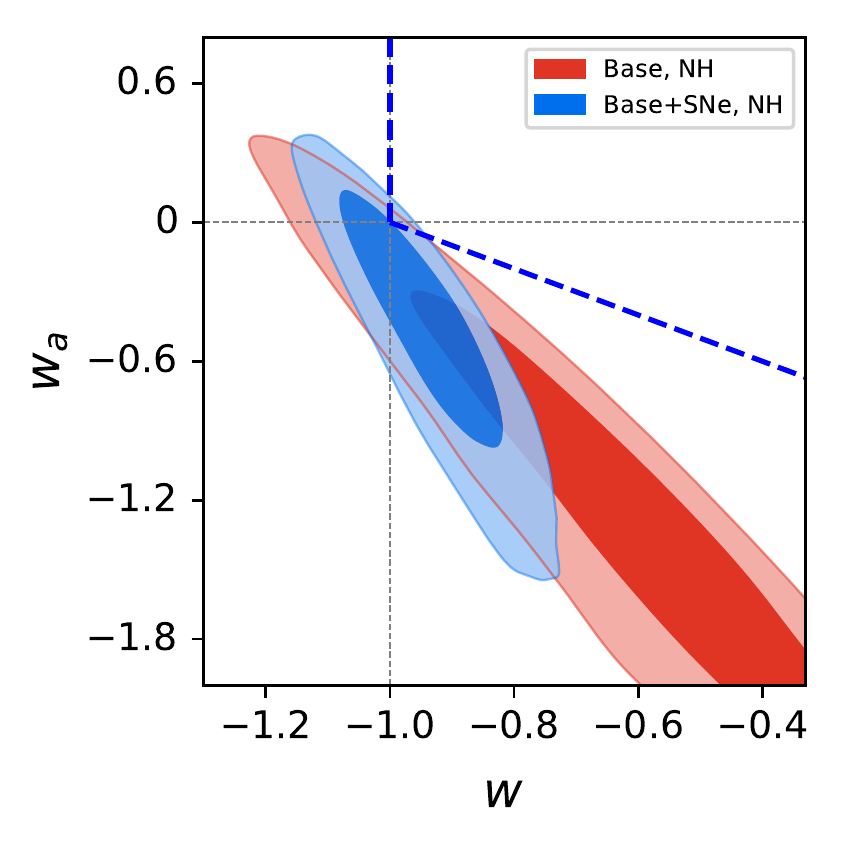}
\hfill
\includegraphics[width=.3\linewidth]{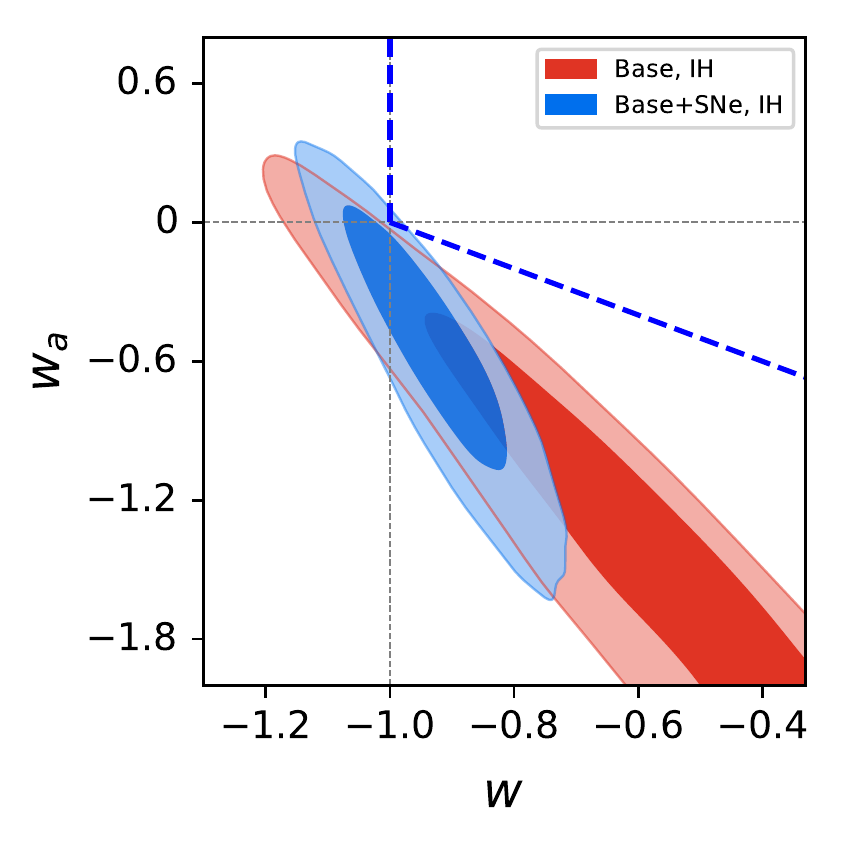}
\caption{\label{fig:4} 68\% and 95\% marginalized contours for $w_0$ vs $w_a$ in the $w_0 w_a\textrm{CDM}+\sum m_{\nu}$ model using Base and Base+SNe datasets, for DH (left panel), NH (middle panel), and IH (right panel). The region at the right of the vertical dashed blue line and above the slanted dashed blue line is the non-phantom DE region.}
\end{figure}

\begin{itemize}
\item iii) \textbf{Results in the $w_0 w_a \textrm{CDM}+\sum m_{\nu}$ model:} In  this model the DE EoS $w(z) = w_0+w_a \frac{z}{1+z}$ is dynamical in nature, but remains close to the value of $w_0+w_a$ for high redshift values, and only changes significantly during the late times. Like the previous model, there is degeneracy between $w(z)$ and $H_0$, and $w(z)$ and $\sum m_{\nu}$. However now we have two parameters $w_0$ and $w_a$ instead of $w$, with 
\begin{equation}
\Omega_{DE}(z) = \Omega_{DE}(0) (1+z)^{3(1+w_0+w_a)} \textrm{exp}\left(-3w_a\frac{z}{1+z}\right).
\end{equation}\label{eq6}
 Any change in $\chi(z_{dec})$ due to $H_0$ can be readily compensated with $w_0$ and $w_a$, and hence, in this model the correlation between $H_0$ and $\sum m_{\nu}$ is very small even with Base data ($R_{H_0,\sum m_{\nu}} =+0.08$ (DH)). Again, $w_0$ and $w_a$ are anti-correlated between themselves, since the change in $\chi(z_{dec})$ due to an increase in $w_0$ can be countered with a decrease in $w_a$. The anti-correlation can be seen clearly in figure \ref{fig:4}. As it can be seen from the figure, neither the Base data nor Base+SNe combination prefers the non-phantom DE region. The data prefers regions where the DE is currently phantom or has been phantom in the past. Since the phantom DE region, $w(z)< -1$ prefers neutrino masses which are larger, the mass bounds are essentially very relaxed in this model, as can be seen from table \ref{tab4}. We have $\sum m_{\nu} < 0.25$ eV (DH), 0.26 eV (NH), 0.28 eV (IH) with Base data, whereas with Base+SNe, we get $\sum m_{\nu} < 0.27$ eV (DH), 0.28 eV (NH), 0.29 eV (IH). The SNe data produces better constraints on the DE parameters.  In figure \ref{fig:4}, another important thing to notice is that as we go from DH to NH to IH, the 2D contours shift away from the non-phantom DE region due to the preference for higher and higher masses, and in the IH case, Base data rejects the non-phantom region at more than 2$\sigma$.
 
\begin{figure}[tbp]
	\centering 
	\includegraphics[width=.45\linewidth]{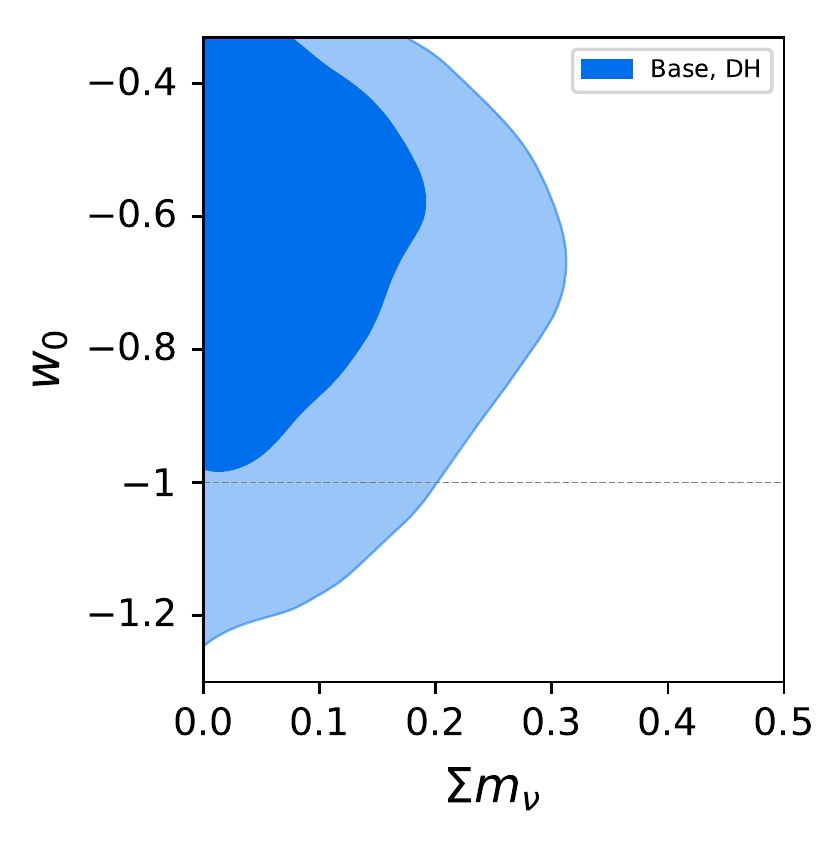}
	\hfill
	\includegraphics[width=.45\linewidth]{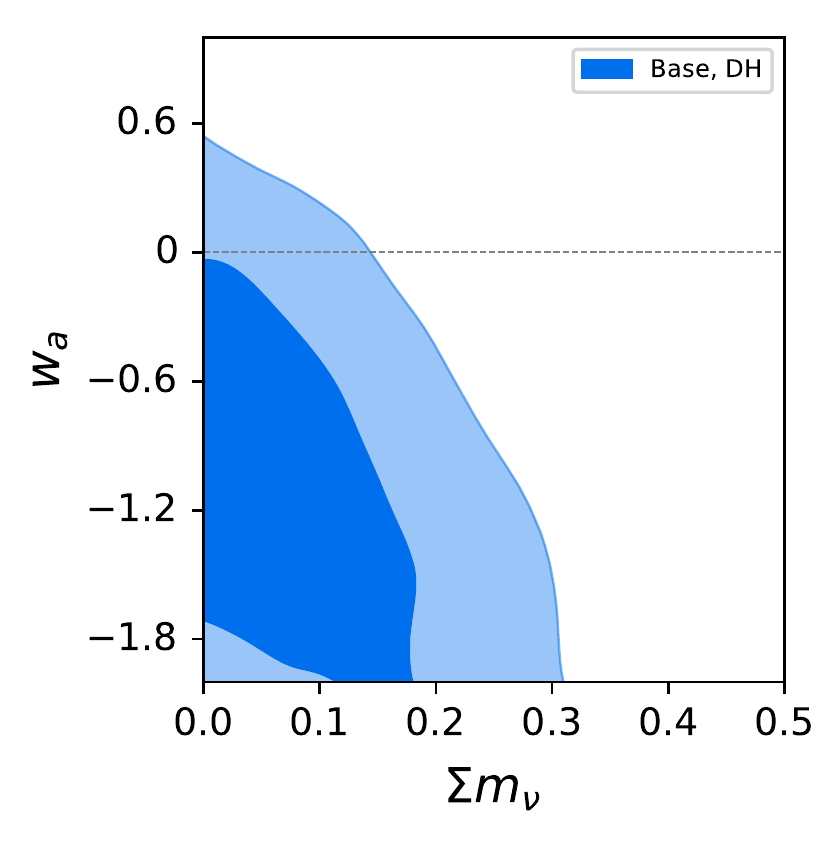}
	\caption{\label{fig:6} 68\% and 95\% marginalized contours for $w_0$ vs $\sum m_\nu$ (eV) (left) and $w_a$ vs $\sum m_\nu$ (eV) (right) in the $w_0 w_a\textrm{CDM}+\sum m_{\nu}$ model using Base dataset, for DH.}
\end{figure} 
 
 We can understand how the individual DE parameters $w_0$ and $w_a$ are affected by $\sum m_\nu$ by looking at figure \ref{fig:6}. For the Base data, the relevant correlation coefficients are: $R_{w_0,\sum m_{\nu}} =-0.004$ (DH), $-0.018$ (NH), $-0.056$ (IH), and $R_{w_a,\sum m_{\nu}} =-0.27$ (DH), $-0.21$ (NH), $-0.17$ (IH). It is clear that $w_a$ is highly anti-correlated with $\sum m_{\nu}$. Since $w(z) = w_0 +w_a z/(1+z)$, given a particular value of $w_0$, it is $w_a$ which determines the late time dynamics of the dark energy equation of state, and thus is more affected by $\sum m_{\nu}$ than $w_0$. Increasing $w_a$ has the effect of increasing $\Omega_{\rm DE}(z)$ at late times ($z\ll1$), as can be seen from eq. \ref{eq6}. This can be partially compensated by decreasing $\sum m_{\nu}$ to decrease $\rho_{\nu} (z)$ at late times to keep $\chi(z_{dec})$ fixed. This can be also compensated by decreasing $w_0$, and thus $w_0$ and $w_a$ are anti-correlated, whereas $w_0$ and $\sum m_{\nu}$ are weakly correlated in the Base data.

\item iv) \textbf{Results in the $w_0 w_a \textrm{CDM}+\sum m_{\nu}$ model with $w(z)\geq -1$ :} This is essentially the same model as in the previous case, but the parameter space is restricted to the non-phantom range only, i.e. $w(z)\geq -1$. This parameter space corresponds to dark energy field theories modelled with a single scalar field, like quintessence \cite{Linder:2007wa}, which cannot cross the phantom barrier (the $w(z)=-1$ line). Due to the degeneracy between $w$ and $\sum m_{\nu}$ (which we have discussed for the last two models) the cosmological data actually prefers smaller and smaller neutrino masses as we go deeper in the non-phantom region in the $w_0-w_a$ parameter space. Thus the bounds on $\sum m_{\nu}$ obtained in this model are even tighter than the $\Lambda\textrm{CDM}+\sum m_{\nu}$ model. This was first noticed in \cite{Vagnozzi:2018jhn}, and subsequently in \cite{Choudhury:2018byy, Choudhury:2018adz}. From table \ref{tab4}, with the base data we find $\sum m_{\nu} < 0.10$ eV (DH), 0.13 eV (NH), 0.16 eV (IH). With the Base+SNe combination, the bounds stand at $\sum m_{\nu} < 0.09$ eV (DH), 0.13 eV (NH), 0.16 eV (IH). These are the tightest bounds on $\sum m_{\nu}$ reported in this paper for the datasets considered. 

\begin{figure}[tbp]
\centering 
\includegraphics[width=.4963\linewidth]{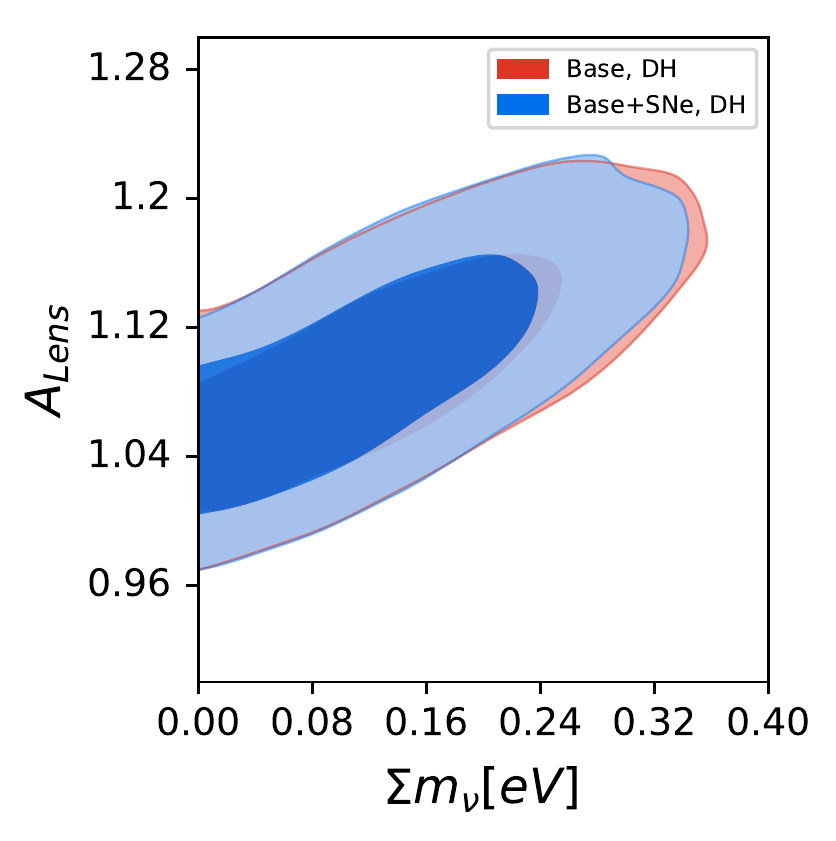}
\hfill
\includegraphics[width=.4963\linewidth]{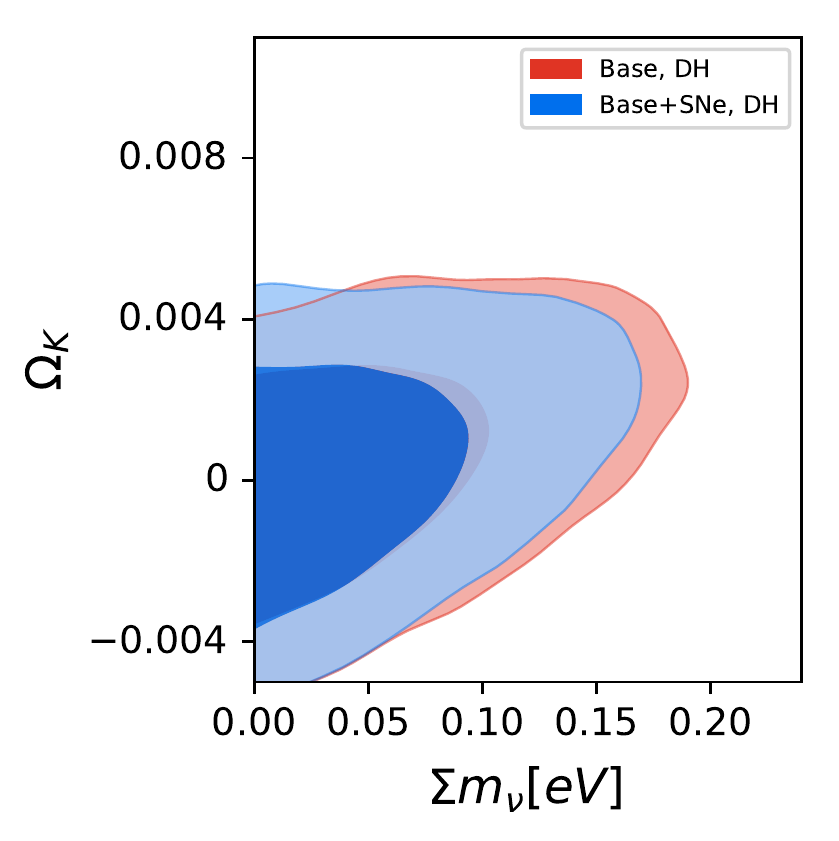}
\caption{\label{fig:5} 68\% and 95\% marginalized contours for $\Omega_k$ vs $\sum m_{\nu}$ in the $\Lambda\textrm{CDM}+\sum m_{\nu}+ \Omega_k$ model on the right, and for $A_{\textrm{Lens}}$ vs $\sum m_{\nu}$ in the $\Lambda\textrm{CDM}+\sum m_{\nu}+A_{\textrm{Lens}}$ model on the left, using Base and Base+SNe datasets and considering degenerate hierarchy.}
\end{figure}

\item v) \textbf{Results in the $\Lambda\textrm{CDM}+\sum m_{\nu}+A_{\textrm{Lens}}$ model :} As mentioned in section \ref{sec2.1}, the $A_{\textrm{Lens}}$ parameter is used to artificially scale the lensing amplitude predicted by the underlying theoretical model. There is a well-known $A_{\textrm{Lens}}$-issue in the CMB anisotropy data which corresponds to preference for $A_{\textrm{Lens}}$ values which are more than 2$\sigma$ away from the theoretical expectation of $A_{\textrm{Lens}}=1$. For instance Planck 2018 TT,TE,EE+lowE data yields an $A_{\textrm{Lens}}=1.180\pm0.065$ (68\%) in the $\Lambda\textrm{CDM}+A_{\textrm{Lens}}$ model, showing a 2.8$\sigma$ tension \cite{Aghanim:2018eyx}. The $\Lambda\textrm{CDM}+A_{\textrm{Lens}}$ also provides significantly better fit to the Planck power spectrum data compare to $\Lambda\textrm{CDM}$. Planck power spectrum constrains $A_{\textrm{Lens}}$ by measuring the smoothing of the CMB acoustic peaks due to the gravitational lensing of the CMB photons from large scale structures. Planck lensing data, however constrains $A_{\textrm{Lens}}$ directly, and adding CMB lensing data to the anisotropy power spectrum usually brings back the $A_{\textrm{Lens}}$ values closer to the theoretically expected result \cite{Ade:2015xua}. We find it true in this model as well. With Base data, for the DH case, we have $A_{\textrm{Lens}} = 1.100^{+0.100}_{-0.096}$ (95\%), which is almost consistent with $A_{\textrm{Lens}}=1$ at 2$\sigma$.  

In this $\Lambda\textrm{CDM}+\sum m_{\nu}+A_{\textrm{Lens}}$ model, $\sum m_{\nu}$ and $A_{\textrm{Lens}}$ are strongly correlated, since increasing $\sum m_{\nu}$ has an effect of reducing lensing induced smearing of the acoustic peaks since increasing $\sum m_{\nu}$ caused increasing suppression to the small scale matter power. We find that $R_{A_{\textrm{Lens}},\sum m_{\nu}} = +$0.68 (DH), $+$0.60 (NH), $+$0.55 (IH) with the Base data. The positive correlation between these two parameters can be seen in figure \ref{fig:5}. This degeneracy causes the bounds on $\sum m_{\nu}$ to relax considerably compared to the $\Lambda\textrm{CDM}+\sum m_{\nu}$ model, as can be seen from table \ref{tab4}. Compared to a 95\% bound of $\sum m_{\nu}<0.12$ eV (DH) with Base data in the minimal  $\Lambda\textrm{CDM}+\sum m_{\nu}$ model, here we get a bound of $\sum m_{\nu}<0.29$ eV with the same data. Rather this $A_{\textrm{Lens}}$-issue actually explains why the neutrino mass bounds are so strong in the $\Lambda\textrm{CDM}+\sum m_{\nu}$ model, especially with the degenerate hierarchy, as a higher $\sum m_{\nu}$ would decrease lensing of the CMB photons. 
From DH to NH to IH, while using same data, the preference for higher $A_{\textrm{Lens}}$ values occurs with higher $\sum m_{\nu}$ values, and the discrepancy with $A_{\textrm{Lens}} = 1$ increases. 

In the absence of varying dark energy EoS, however, $H_0$ and $\sum m_{\nu}$ are strongly correlated in this model. The model also produces lower $\sigma_8$ and $S_8$ values compared to other models considered in the paper, leading to a decrease in the $\sigma_8$ tension present between Planck and cosmic shear experiments like CFHTLenS \cite{Erben:2012zw}, KiDS-450 \cite{Hildebrandt:2016iqg} etc. 

\item v) \textbf{Results in the $\Lambda\textrm{CDM}+\sum m_{\nu}+\Omega_{\textrm{k}}$ model :} In this model we consider the possibility of a non-flat background geometry of the universe. $\Omega_{\textrm{k}}$ and $\sum m_{\nu}$ are correlated positively and strongly with both Base and Base+SNe combinations, as can be seen from the right panel of figure \ref{fig:5}. This causes the bounds on $\sum m_{\nu}$ to degrade compared to the  $\Lambda\textrm{CDM}+\sum m_{\nu}$ model. With Base data we have: $\sum m_{\nu} < 0.15$ eV (DH), $\sum m_{\nu} < 0.17$ eV (DH), $\sum m_{\nu} < 0.20$ eV (IH). The origin the degeneracy again lies in the tightly constrained angular diameter distance to the last scattering surface, i.e. $D_{A}(z_{dec}) \equiv  \textrm{sin}\left(\sqrt{K\chi(z_{\rm dec}}\right)/\sqrt{K}$ (where $K = -\Omega_k H_0^2$ is the curvature), leading to a three way geometric degeneracy between $H_0$, $\Omega_k$ and $\sum m_{\nu}$ \cite{Howlett:2012mh} (note that when $\Omega_k=0$, $D_{A}(z) = \chi(z)$). In this model, the expression for $H(z)$ is given by,
\begin{equation}
H(z) = \sqrt{\omega_{\gamma}(1+z)^4 + (\omega_c + \omega_b)(1+z)^3 + \omega_{\Lambda}+\omega_k (1+z)^2 + \rho_{\nu}(z)h^2/\rho_{cr,0}}
\end{equation}.

It is to be noted however, that the Planck data actually prefers values of $\Omega_{\textrm{k}}<0$ \cite{Aghanim:2018eyx}, since closed universe models produce more lensing amplitude compared to a flat universe. However, inclusion of lensing data usually brings the parameters back closer to a flat universe. BAO data helps in partially breaking the three way degeneracy by constraining $H_0$. We find that with Base data, in the DH case $\Omega_{\textrm{k}} = 0.0004\pm0.0020$, which is perfectly consistent with $\Omega_{\textrm{k}}=0$. The correlation coefficients are $R_{\Omega_k, \sum m_{\nu}} = +0.41$,  $R_{H_0, \sum m_{\nu}} = -0.18$, and $R_{\Omega_k, H_0} = +0.60$ for DH, i.e. the three parameters still remain considerably correlated with each other. From DH to NH to IH, preference for higher neutrino masses leads to preference for higher values of $\Omega_k$, although with the Base data $\Omega_{\textrm{k}} = 0$ is included within 2$\sigma$ ranges of $\Omega_k$, for any of the hierarchies. 

\end{itemize}

\textbf{Akaike information criterion (AIC).} Here we compare the goodness of fit of degenerate, normal and inverted hierarchy scenarios in the extended models that we have studied in this section. As in previous section, we use AIC as a measure of goodness of fit. AIC for the degenerate, normal, and inverted hierarchy are denoted as $\textrm{AIC}_{\textrm{DH}}$, $\textrm{AIC}_{\textrm{NH}}$, and $\textrm{AIC}_{\textrm{IH}}$ respectively. Since the DH, NH and IH cases both have the same number of parameters, $\Delta \textrm{AIC} = \textrm{AIC} - \textrm{AIC}_{\textrm{IH}} = \Delta \chi^2$, where $\Delta \chi^2$ is the $\chi^2$ difference between the $\chi^2$ value for a particular hierarchy (DH, NH, or IH) and the IH case, at the best-fit points. The $\Delta \textrm{AIC}$ values for each of the extended models has been listed in table \ref{tab5}, for the Base and Base+SNe datasets. We find that in most scenarios $\Delta \textrm{AIC} < 0$, i.e. the fit due to DH/NH is better than IH, whereas in a few cases $\Delta \textrm{AIC} > 0$. But in none of the models do we see any statistically significant difference between NH and IH. 

\begin{table*}\centering
	\ra{1.3}
	
	\resizebox{\textwidth}{!}{\begin{tabular}{@{}rrrrcrrr@{}}\toprule
			& \multicolumn{3}{c}{Base} & \phantom{abc} & \multicolumn{3}{c}{Base+SNe}\\
			\cmidrule{2-4} \cmidrule{6-8}
			& DH & NH & IH && DH & NH & IH\\ \midrule
	$\Lambda\textrm{CDM}+\sum m_{\nu}+r$  	& $-5.32$ & $-1.39$ & 0 && $-4.46$ & $-3.21$ & 0 \\
	
	$w\textrm{CDM}+\sum m_{\nu}$      	& $-1.66$ & $-1.59$ & 0 && $-0.84$ & $-0.99$ & 0 \\

	$w_0 w_a \textrm{CDM}+\sum m_{\nu}$   	& $+0.75$ & $+2.08$ & 0 && $-0.80$ & $-0.41$ & 0 \\

	$w_0 w_a \textrm{CDM}+\sum m_{\nu}$ with $w(z)\geq -1$ 	& $-2.71$ & $-0.69$ & 0 && $-4.67$ &$-1.31$  & 0\\
	
	$\Lambda\textrm{CDM}+\sum m_{\nu}+A_{\textrm{Lens}}$ 	& $+0.88$ & $+0.20$ & 0 && $+0.13$ &  $+0.92$ & 0 \\
	
	$\Lambda\textrm{CDM}+\sum m_{\nu}+\Omega_{\textrm{k}}$ & $-3.56$ & $-1.96$ & 0 && $-2.74$ & $-1.22$  & 0 \\ 
	
		\bottomrule
    \end{tabular}}
    \caption{ The values of $\Delta \chi^2 = \Delta \textrm{AIC} = \chi^2 - \chi^2_{\rm IH}$ for various extended models studied in this paper, with Base and Base+SNe dataset combinations. The $\chi^2$ differences are calculated at best-fit points.}\label{tab5}
\end{table*}

\section{Bayesian Model Comparison}\label{sec5}

While in the previous section we had chosen a frequentist measure (i.e. $\Delta\rm AIC$) to compare NH and IH, in this section we take a Bayesian approach. Our method closely follows \cite{Hannestad:2016fog}, and that of \cite{Blennow:2013kga,Hall:2012kg}. Consider that any two models, $M_1$ and $M_2$, with parameter vectors $\theta_1$ and $\theta_2$, are being tested against the same data $D$. Then, the posterior probability of a particular model $M_i$ ($i\equiv 1,2$) given the data $D$ is,

\begin{equation}
p_i \equiv P(M_i|D) = \frac{\pi_i P(D|M_i)}{P(D)}, 
\end{equation}\label{eq7}
where $P(D)$ is the prior probability for the data, and $\pi_i\equiv P(M_i)$ is our prior degree of belief in model $M_i$. Since we are dealing with probabilities, if $M_1$ and $M_2$ are the only two possible models for a particular scenario(for instance, if $M_1$ and $M_2$ represent NH and IH and we know these are the only two possible mass orderings), we can write $\pi_1 + \pi_2 =1$.
$P(D|M_i)$ is the Bayesian evidence or the marginal likelihood, and is calculated by marginalizing product of the likelihood $P(D|\theta_i,M_i)$ and the parameter prior $P(\theta_i|M_i)$ over all parameters:
\begin{equation}
P(D|M_i) = \int d\theta_i P(D|\theta_i,M_i) P(\theta_i|M_i). 
\end{equation} 

Again, as we can take the $p_1 + p_2 =1$, we can rewrite eq.~\ref{eq7} as,

\begin{equation}
p_i \equiv P(M_i|D) = \frac{\pi_i P(D|M_i)}{\Sigma_i\pi_i P(D|M_i)}.
\end{equation}

As indicated before, we shall treat $M_1$ and $M_2$ as NH and IH, and from now on we shall use the notation $p_{\rm NH}$ and $p_{\rm IH}$ for their posterior probabilities. One can now define the quantity Odds(NH:IH) as
\begin{equation}
\mathrm{Odds(NH:IH)} \equiv \frac{p_{\rm NH}}{p_{\rm IH}}
\end{equation}
which is simply the ratio of posterior probabilities \cite{Blennow:2013kga}. If no apriori knowledge of $\pi_i$ exists, it is customary to take all the $\pi_i$ to be equal, i.e. $\pi_{\rm NH} = \pi_{\rm IH} = 0.5$. In that case, Odds(NH:IH) essentially becomes equal to the Bayes' factor $B_{\rm NH/IH}$, which is defined as the ratio of evidences in the NH and IH case. 

However, in case of neutrino mass hierarchy, one can assign $\pi_i$ from the Bayesian analysis of neutrino oscillations data. From the Bayesian analysis of neutrino oscillations data in \cite{deSalas:2018bym}, we see that  $ln(B_{\rm NH/IH}) = 6.5$ (taking only the central value), which points to a posterior $\textrm{Odds(NH:IH)} = 665:1$. If we take this as our prior belief before doing any analysis with cosmological data, we get $\pi_{\rm NH} = 0.9985, \pi_{\rm IH} = 0.0015$. We have used both this and the previous $\pi_{\rm NH} = \pi_{\rm IH} = 0.5$ case while calculating the Odds(NH:IH).

Alternatively, following \cite{Hannestad:2016fog, Vagnozzi:2017ovm}, one can also consider the confidence limit at which IH is disfavored, as  

\begin{equation}
\mathrm{CL_{IH}} \equiv 1- p_{\rm IH}.
\end{equation}

 Odds(NH:IH) and CL$_{\rm IH}$ and strength of evidence as per Jeffrey’s scale \cite{Jeffreys} has been provided in table \ref{tab6}, where we have used only the mean values of the estimated Bayesian evidence (by CosmoChord \cite{Handley}) to compute the numerical quantities. It can be seen from table~\ref{tab6} that the analyses without the input from neutrino oscillations data on $\pi$ yield only a ``Inconclusive'' evidence for the normal neutrino mass hierarchy, which agrees with with our small $\Delta \rm AIC$ values from previous section. On the other hand, all the evidences are ``Strong'' with $\pi_{\rm NH} = 0.9985, \pi_{\rm IH} = 0.0015$, i.e. the preference for normal mass hierarchy is mostly driven by the oscillations data.

\begin{table*}\centering
	\ra{1.3}
	
	\resizebox{\textwidth}{!}{\begin{tabular}{@{}rrrccrrc@{}}\toprule
			& \multicolumn{3}{c}{$\pi_{\rm NH}=\pi_{\rm IH}=0.5$} & \phantom{abc} & \multicolumn{3}{c}{$\pi_{\rm NH}=0.9985$,$\pi_{\rm IH}=0.0015$}\\
			\cmidrule{2-4} \cmidrule{6-8}
			& Odds(NH:IH) & CL$_{\rm IH}$ & Strength of Evidence && Odds(NH:IH) & CL$_{\rm IH}$ & Strength of Evidence\\ \midrule
			$\Lambda\textrm{CDM}+\sum m_{\nu}$  	& 1.43:1 & 58.8\% & Inconclusive && 950:1 & 99.895\% & Strong \\
			$\Lambda\textrm{CDM}+\sum m_{\nu}+r$  	& 0.93:1 & 48.1\% & Inconclusive && 617:1 & 99.838\% & Strong \\
			
			$w\textrm{CDM}+\sum m_{\nu}$      	& 1.3:1 & 56.6\% & Inconclusive && 867:1 & 99.885\% & Strong\\
			
			$w_0 w_a \textrm{CDM}+\sum m_{\nu}$   	& 1.24:1 & 55.4\% & Inconclusive && 827:1 & 99.879\% & Strong \\
			
			$w_0 w_a \textrm{CDM}+\sum m_{\nu}$ with $w(z)\geq -1$ 	& 2:1 & 66.6\% & Inconclusive && 1329:1 & 99.925\%  & Strong\\
			
			$\Lambda\textrm{CDM}+\sum m_{\nu}+A_{\textrm{Lens}}$ 	& 0.75:1 & 42.9\% & Inconclusive && 500:1 &  99.800\% & Strong \\
			
			$\Lambda\textrm{CDM}+\sum m_{\nu}+\Omega_{\textrm{k}}$ & 2.3:1 & 69.7\% & Inconclusive && 1530:1 & 99.935\%  & Strong \\ 
			
			\bottomrule
	\end{tabular}}
	\caption{ The values of Odds(NH:IH) and CL$_{\rm IH}$ and strength of evidence for various cosmological models studied in this paper, with Base dataset.}\label{tab6}
\end{table*}

\section{Discussion and conclusions}\label{sec4}
Presently the upper bounds on the sum of 3 active neutrino masses, $\sum m_{\nu}$ from analyses of cosmological data in the backdrop of $\Lambda\textrm{CDM}+\sum m_{\nu}$ model are bordering on the minimum sum of neutrino masses required by the inverted hierarchy, which is around 0.1 eV. However, these analyses are usually done with the assumption of 3 degenerate neutrino masses, but terrestrial neutrino oscillation experiments have confirmed that the three neutrino masses are not equal. Thus in this paper we update the bounds on $\sum m_{\nu}$ from latest publicly available cosmological data while explicitly considering particular neutrino mass hierarchies using the results on mass-squared splittings from a global analysis of neutrino oscillations data, NuFit 4.0 \cite{Esteban:2018azc}. For implementing the normal and inverted hierarchy scenarios, we use the mass of the lightest mass eigenstate, denoted with $m_0$, as the varying parameter, and use the mass-squared splittings from NuFit 4.0 to determine the other masses in a particular hierarchy, and thus use the total mass sum, i.e. $\sum m_{\nu}$, as a derived parameter. This approach puts some implicit priors on the neutrino mass sum: $\sum m_{\nu} \geq 0.06$ eV for the normal hierarchy (NH) case,  $\sum m_{\nu} \geq 0.10$ eV for the inverted hierarchy (IH) case.   

In the minimal $\Lambda\textrm{CDM}+\sum m_{\nu}$ model with Planck 2018 TT,TE,EE, lowE, lensing, and the latest BAO data from various galaxy surveys, we find that at 95\% C.L. $\sum m_{\nu}<0.12$ eV in the case of degenerate mass approximation. We call this dataset ``Base."  This is similar to the bound of $\sum m_{\nu}<0.12$ eV quoted by the Planck 2018 collaboration using the same data, in the same model. We also find that in the same model we have  $\sum m_{\nu}<0.15$ eV in case of NH and $\sum m_{\nu}<0.17$ eV in case of IH; i.e., the bounds vary significantly across the different mass orderings. The main reason for these differences in the bounds is the implicit priors on $\sum m_{\nu}$ when we assume a particular hierarchy. The case of degenerate neutrino masses with a prior $\sum m_{\nu}\geq 0.06$ eV or $\sum m_{\nu}\geq 0.10$ eV produces almost the same bounds as the NH or IH case with the lightest mass ($m_0$) parametrization we have used in the paper. The NH and IH case produce bounds which are only slightly stronger than the degenerate case with priors of $\sum m_{\nu}\geq 0.06$ eV and $\sum m_{\nu}\geq 0.10$ eV respectively, which can be attributed to the fact that with the lightest mass ($m_0$) parametrization, the implicit prior on $\sum m_{\nu}$ is not completely flat (i.e. rises at the low values of $m_0$). Also, we find that the normal hierarchy is very mildly preferred to the inverted: $\Delta \chi^2 \equiv \chi^2_{\textrm{NH}}- \chi^2_{\textrm{IH}} = -0.95$ (best-fit). We also study this model against another dataset combination: Base+SNe. For Base and Base+SNe datasets, the $\chi^2$ differences between the NH and IH case remain statistically mild, but the bounds on $\sum m_{\nu}$ across the three different mass orderings vary considerably. 

In this paper, we also provide bounds on $\sum m_{\nu}$ considering different hierarchies in various extended models: $\Lambda\textrm{CDM}+\sum m_{\nu}+r$, $w\textrm{CDM}+\sum m_{\nu}$, $w_0 w_a \textrm{CDM}+\sum m_{\nu}$, $w_0 w_a \textrm{CDM}+\sum m_{\nu}$ model with $w(z)\geq -1$, $\Lambda \textrm{CDM} + \sum m_{\nu} + A_{\textrm{Lens}}$, and $\Lambda \textrm{CDM} + \sum m_{\nu} + \Omega_k$. Here $r$ is the tensor-to-scalar ratio, $w$ is the dark energy equation of state (DE EoS) parameter, $w_0$ and $w_a$ again parametrize the DE EoS but in a dynamical manner: $w(z) = w_0 + w_a z/(1+z)$ (CPL parametrization), $A_{\textrm{Lens}}$ is the parameter for artificial scaling of the lensing amplitude, and $\Omega_k$ is the curvature energy density. We used the Base and Base+SNe dataset combinations to constrain these models. 

Consistent with other studies (see e.g.\ \cite{DiValentino:2019dzu} for a very recent example) we found that in some cases the formal bound $\sum m_\nu$ could be loosened by up to factor of two.
However, in none of the extended models could we find any statistically significant difference in the quality of fit to the data between NH and IH, i.e. the current cosmological data is not sufficiently strong to demarcate the two hierarchies. Apart from checking $\chi^2$ differences between NH and IH, we also confirmed the same using computation of Bayesian evidence. The results are provided in section {\ref{sec5}}. Starting with equal apriori beliefs on NH and IH scenarios, we find no conclusive evidence for the normal hierarchy over inverted hierarchy in any of the cosmological models with the Base dataset. This finding is consistent with previous work showing that if the actual value of $\sum m_{\nu}$ is sufficiently less than 0.1 eV (i.e. the neutrino masses follow NH) a formal sensitivity to $\sum m_\nu$ of 0.01-0.02 eV is required to guarantee a conclusive distinction between the two hierarchies \cite{Hannestad:2016fog,Mahony:2019fyb}. However, if the actual value of $\sum m_{\nu} >$ 0.1 eV, then we will not be able to determine the hierarchy, due to the fact that even the most optimistic future cosmological measurements won't be able to differentiate among individual neutrino masses \cite{Archidiacono:2020dvx}. Hence, for a direct detection of the neutrino mass hierarchy, we have to depend on future terrestrial neutrino oscillation experiments like DUNE \cite{Acciarri:2015uup} and JUNO \cite{An:2015jdp}. 

\section*{Acknowledgements}
The authors thank the anonymous referee immensely, for the suggestions on improving the manuscript. for SRC thanks the computing facilities at HRI (\url{http://www.hri.res.in/cluster/}) and at Aarhus University (\url{http://www.cscaa.dk/}), and the warm hospitality at Aarhus University during the completion of the project. SRC would also like to thank the Department of Atomic Energy (DAE) Neutrino Project of HRI. STH is supported by a grant from the Villum Foundation. 

\bibliography{paper}

\providecommand{\href}[2]{#2}\begingroup\raggedright\begin{thebibliography}{100}

\bibitem{Abe:2013hdq}
{\scshape T2K} collaboration, K.~Abe et~al., \emph{{Observation of Electron
  Neutrino Appearance in a Muon Neutrino Beam}},
  \href{https://doi.org/10.1103/PhysRevLett.112.061802}{\emph{Phys. Rev. Lett.}
  {\bfseries 112} (2014) 061802}
  [\href{https://arxiv.org/abs/1311.4750}{{\ttfamily 1311.4750}}].

\bibitem{Ahn:2012nd}
{\scshape RENO} collaboration, J.~K. Ahn et~al., \emph{{Observation of Reactor
  Electron Antineutrino Disappearance in the RENO Experiment}},
  \href{https://doi.org/10.1103/PhysRevLett.108.191802}{\emph{Phys. Rev. Lett.}
  {\bfseries 108} (2012) 191802}
  [\href{https://arxiv.org/abs/1204.0626}{{\ttfamily 1204.0626}}].

\bibitem{Abe:2012tg}
{\scshape Double Chooz} collaboration, Y.~Abe et~al., \emph{{Reactor electron
  antineutrino disappearance in the Double Chooz experiment}},
  \href{https://doi.org/10.1103/PhysRevD.86.052008}{\emph{Phys. Rev.}
  {\bfseries D86} (2012) 052008}
  [\href{https://arxiv.org/abs/1207.6632}{{\ttfamily 1207.6632}}].

\bibitem{An:2012eh}
{\scshape Daya Bay} collaboration, F.~P. An et~al., \emph{{Observation of
  electron-antineutrino disappearance at Daya Bay}},
  \href{https://doi.org/10.1103/PhysRevLett.108.171803}{\emph{Phys. Rev. Lett.}
  {\bfseries 108} (2012) 171803}
  [\href{https://arxiv.org/abs/1203.1669}{{\ttfamily 1203.1669}}].

\bibitem{Araki:2004mb}
{\scshape KamLAND} collaboration, T.~Araki et~al., \emph{{Measurement of
  neutrino oscillation with KamLAND: Evidence of spectral distortion}},
  \href{https://doi.org/10.1103/PhysRevLett.94.081801}{\emph{Phys. Rev. Lett.}
  {\bfseries 94} (2005) 081801}
  [\href{https://arxiv.org/abs/hep-ex/0406035}{{\ttfamily hep-ex/0406035}}].

\bibitem{Adamson:2008zt}
{\scshape MINOS} collaboration, P.~Adamson et~al., \emph{{Measurement of
  Neutrino Oscillations with the MINOS Detectors in the NuMI Beam}},
  \href{https://doi.org/10.1103/PhysRevLett.101.131802}{\emph{Phys. Rev. Lett.}
  {\bfseries 101} (2008) 131802}
  [\href{https://arxiv.org/abs/0806.2237}{{\ttfamily 0806.2237}}].

\bibitem{Fukuda:1998mi}
{\scshape Super-Kamiokande} collaboration, Y.~Fukuda et~al., \emph{{Evidence
  for oscillation of atmospheric neutrinos}},
  \href{https://doi.org/10.1103/PhysRevLett.81.1562}{\emph{Phys. Rev. Lett.}
  {\bfseries 81} (1998) 1562}
  [\href{https://arxiv.org/abs/hep-ex/9807003}{{\ttfamily hep-ex/9807003}}].

\bibitem{Ahmad:2002jz}
{\scshape SNO} collaboration, Q.~R. Ahmad et~al., \emph{{Direct evidence for
  neutrino flavor transformation from neutral current interactions in the
  Sudbury Neutrino Observatory}},
  \href{https://doi.org/10.1103/PhysRevLett.89.011301}{\emph{Phys. Rev. Lett.}
  {\bfseries 89} (2002) 011301}
  [\href{https://arxiv.org/abs/nucl-ex/0204008}{{\ttfamily nucl-ex/0204008}}].

\bibitem{Esteban:2018azc}
I.~Esteban, M.~C. Gonzalez-Garcia, A.~Hernandez-Cabezudo, M.~Maltoni and
  T.~Schwetz, \emph{{Global analysis of three-flavour neutrino oscillations:
  synergies and tensions in the determination of $\theta_23, \delta_CP$, and
  the mass ordering}},
  \href{https://doi.org/10.1007/JHEP01(2019)106}{\emph{JHEP} {\bfseries 01}
  (2019) 106} [\href{https://arxiv.org/abs/1811.05487}{{\ttfamily
  1811.05487}}].

\bibitem{Forero:2014bxa}
D.~V. Forero, M.~Tortola and J.~W.~F. Valle, \emph{{Neutrino oscillations
  refitted}}, \href{https://doi.org/10.1103/PhysRevD.90.093006}{\emph{Phys.
  Rev.} {\bfseries D90} (2014) 093006}
  [\href{https://arxiv.org/abs/1405.7540}{{\ttfamily 1405.7540}}].

\bibitem{Gonzalez-Garcia:2014bfa}
M.~C. Gonzalez-Garcia, M.~Maltoni and T.~Schwetz, \emph{{Updated fit to three
  neutrino mixing: status of leptonic CP violation}},
  \href{https://doi.org/10.1007/JHEP11(2014)052}{\emph{JHEP} {\bfseries 11}
  (2014) 052} [\href{https://arxiv.org/abs/1409.5439}{{\ttfamily 1409.5439}}].

\bibitem{Esteban:2016qun}
I.~Esteban, M.~C. Gonzalez-Garcia, M.~Maltoni, I.~Martinez-Soler and
  T.~Schwetz, \emph{{Updated fit to three neutrino mixing: exploring the
  accelerator-reactor complementarity}},
  \href{https://doi.org/10.1007/JHEP01(2017)087}{\emph{JHEP} {\bfseries 01}
  (2017) 087} [\href{https://arxiv.org/abs/1611.01514}{{\ttfamily
  1611.01514}}].

\bibitem{Capozzi:2016rtj}
F.~Capozzi, E.~Lisi, A.~Marrone, D.~Montanino and A.~Palazzo, \emph{{Neutrino
  masses and mixings: Status of known and unknown $3\nu$ parameters}},
  \href{https://doi.org/10.1016/j.nuclphysb.2016.02.016}{\emph{Nucl. Phys.}
  {\bfseries B908} (2016) 218}
  [\href{https://arxiv.org/abs/1601.07777}{{\ttfamily 1601.07777}}].

\bibitem{Capozzi:2017ipn}
F.~Capozzi, E.~Di~Valentino, E.~Lisi, A.~Marrone, A.~Melchiorri and A.~Palazzo,
  \emph{{Global constraints on absolute neutrino masses and their ordering}},
  \href{https://doi.org/10.1103/PhysRevD.95.096014}{\emph{Phys. Rev.}
  {\bfseries D95} (2017) 096014}
  [\href{https://arxiv.org/abs/1703.04471}{{\ttfamily 1703.04471}}].

\bibitem{Caldwell:2017mqu}
A.~Caldwell, A.~Merle, O.~Schulz and M.~Totzauer, \emph{{Global Bayesian
  analysis of neutrino mass data}},
  \href{https://doi.org/10.1103/PhysRevD.96.073001}{\emph{Phys. Rev.}
  {\bfseries D96} (2017) 073001}
  [\href{https://arxiv.org/abs/1705.01945}{{\ttfamily 1705.01945}}].

\bibitem{Lesgourgues:2006nd}
J.~Lesgourgues and S.~Pastor, \emph{{Massive neutrinos and cosmology}},
  \href{https://doi.org/10.1016/j.physrep.2006.04.001}{\emph{Phys. Rept.}
  {\bfseries 429} (2006) 307}
  [\href{https://arxiv.org/abs/astro-ph/0603494}{{\ttfamily
  astro-ph/0603494}}].

\bibitem{Wong:2011ip}
Y.~Y.~Y. Wong, \emph{{Neutrino mass in cosmology: status and prospects}},
  \href{https://doi.org/10.1146/annurev-nucl-102010-130252}{\emph{Ann. Rev.
  Nucl. Part. Sci.} {\bfseries 61} (2011) 69}
  [\href{https://arxiv.org/abs/1111.1436}{{\ttfamily 1111.1436}}].

\bibitem{Lesgourgues:2012uu}
J.~Lesgourgues and S.~Pastor, \emph{{Neutrino mass from Cosmology}},
  \href{https://doi.org/10.1155/2012/608515}{\emph{Adv. High Energy Phys.}
  {\bfseries 2012} (2012) 608515}
  [\href{https://arxiv.org/abs/1212.6154}{{\ttfamily 1212.6154}}].

\bibitem{Abazajian:2013oma}
{\scshape Topical Conveners: K.N. Abazajian, J.E. Carlstrom, A.T. Lee}
  collaboration, K.~N. Abazajian et~al., \emph{{Neutrino Physics from the
  Cosmic Microwave Background and Large Scale Structure}},
  \href{https://doi.org/10.1016/j.astropartphys.2014.05.014}{\emph{Astropart.
  Phys.} {\bfseries 63} (2015) 66}
  [\href{https://arxiv.org/abs/1309.5383}{{\ttfamily 1309.5383}}].

\bibitem{Lesgourgues:2014zoa}
J.~Lesgourgues and S.~Pastor, \emph{{Neutrino cosmology and Planck}},
  \href{https://doi.org/10.1088/1367-2630/16/6/065002}{\emph{New J. Phys.}
  {\bfseries 16} (2014) 065002}
  [\href{https://arxiv.org/abs/1404.1740}{{\ttfamily 1404.1740}}].

\bibitem{Archidiacono:2016lnv}
M.~Archidiacono, T.~Brinckmann, J.~Lesgourgues and V.~Poulin, \emph{{Physical
  effects involved in the measurements of neutrino masses with future
  cosmological data}},
  \href{https://doi.org/10.1088/1475-7516/2017/02/052}{\emph{JCAP} {\bfseries
  1702} (2017) 052} [\href{https://arxiv.org/abs/1610.09852}{{\ttfamily
  1610.09852}}].

\bibitem{Lattanzi:2017ubx}
M.~Lattanzi and M.~Gerbino, \emph{{Status of neutrino properties and future
  prospects - Cosmological and astrophysical constraints}},
  \href{https://doi.org/10.3389/fphy.2017.00070}{\emph{Front.in Phys.}
  {\bfseries 5} (2018) 70} [\href{https://arxiv.org/abs/1712.07109}{{\ttfamily
  1712.07109}}].

\bibitem{Choudhury:2018byy}
S.~Roy~Choudhury and S.~Choubey, \emph{{Updated Bounds on Sum of Neutrino
  Masses in Various Cosmological Scenarios}},
  \href{https://doi.org/10.1088/1475-7516/2018/09/017}{\emph{JCAP} {\bfseries
  1809} (2018) 017} [\href{https://arxiv.org/abs/1806.10832}{{\ttfamily
  1806.10832}}].

\bibitem{Aghanim:2018eyx}
{\scshape Planck} collaboration, N.~Aghanim et~al., \emph{{Planck 2018 results.
  VI. Cosmological parameters}},
  \href{https://arxiv.org/abs/1807.06209}{{\ttfamily 1807.06209}}.

\bibitem{Vagnozzi:2017ovm}
S.~Vagnozzi, E.~Giusarma, O.~Mena, K.~Freese, M.~Gerbino, S.~Ho et~al.,
  \emph{{Unveiling $\nu$ secrets with cosmological data: neutrino masses and
  mass hierarchy}},
  \href{https://doi.org/10.1103/PhysRevD.96.123503}{\emph{Phys. Rev.}
  {\bfseries D96} (2017) 123503}
  [\href{https://arxiv.org/abs/1701.08172}{{\ttfamily 1701.08172}}].

\bibitem{Palanque-Delabrouille:2015pga}
N.~Palanque-Delabrouille et~al., \emph{{Neutrino masses and cosmology with
  Lyman-alpha forest power spectrum}},
  \href{https://doi.org/10.1088/1475-7516/2015/11/011}{\emph{JCAP} {\bfseries
  1511} (2015) 011} [\href{https://arxiv.org/abs/1506.05976}{{\ttfamily
  1506.05976}}].

\bibitem{DiValentino:2015wba}
E.~Di~Valentino, E.~Giusarma, M.~Lattanzi, O.~Mena, A.~Melchiorri and J.~Silk,
  \emph{{Cosmological Axion and neutrino mass constraints from Planck 2015
  temperature and polarization data}},
  \href{https://doi.org/10.1016/j.physletb.2015.11.025}{\emph{Phys. Lett.}
  {\bfseries B752} (2016) 182}
  [\href{https://arxiv.org/abs/1507.08665}{{\ttfamily 1507.08665}}].

\bibitem{Cuesta:2015iho}
A.~J. Cuesta, V.~Niro and L.~Verde, \emph{{Neutrino mass limits: robust
  information from the power spectrum of galaxy surveys}},
  \href{https://doi.org/10.1016/j.dark.2016.04.005}{\emph{Phys. Dark Univ.}
  {\bfseries 13} (2016) 77} [\href{https://arxiv.org/abs/1511.05983}{{\ttfamily
  1511.05983}}].

\bibitem{Huang:2015wrx}
Q.-G. Huang, K.~Wang and S.~Wang, \emph{{Constraints on the neutrino mass and
  mass hierarchy from cosmological observations}},
  \href{https://doi.org/10.1140/epjc/s10052-016-4334-z}{\emph{Eur. Phys. J.}
  {\bfseries C76} (2016) 489}
  [\href{https://arxiv.org/abs/1512.05899}{{\ttfamily 1512.05899}}].

\bibitem{Moresco:2016nqq}
M.~Moresco, R.~Jimenez, L.~Verde, A.~Cimatti, L.~Pozzetti, C.~Maraston et~al.,
  \emph{{Constraining the time evolution of dark energy, curvature and neutrino
  properties with cosmic chronometers}},
  \href{https://doi.org/10.1088/1475-7516/2016/12/039}{\emph{JCAP} {\bfseries
  1612} (2016) 039} [\href{https://arxiv.org/abs/1604.00183}{{\ttfamily
  1604.00183}}].

\bibitem{Giusarma:2016phn}
E.~Giusarma, M.~Gerbino, O.~Mena, S.~Vagnozzi, S.~Ho and K.~Freese,
  \emph{{Improvement of cosmological neutrino mass bounds}},
  \href{https://doi.org/10.1103/PhysRevD.94.083522}{\emph{Phys. Rev.}
  {\bfseries D94} (2016) 083522}
  [\href{https://arxiv.org/abs/1605.04320}{{\ttfamily 1605.04320}}].

\bibitem{Couchot:2017pvz}
F.~Couchot, S.~Henrot-Versillé, O.~Perdereau, S.~Plaszczynski,
  B.~Rouillé~d'Orfeuil, M.~Spinelli et~al., \emph{{Cosmological constraints on
  the neutrino mass including systematic uncertainties}},
  \href{https://doi.org/10.1051/0004-6361/201730927}{\emph{Astron. Astrophys.}
  {\bfseries 606} (2017) A104}
  [\href{https://arxiv.org/abs/1703.10829}{{\ttfamily 1703.10829}}].

\bibitem{Doux:2017tsv}
C.~Doux, M.~Penna-Lima, S.~D.~P. Vitenti, J.~Tréguer, E.~Aubourg and K.~Ganga,
  \emph{{Cosmological constraints from a joint analysis of cosmic microwave
  background and spectroscopic tracers of the large-scale structure}},
  \href{https://doi.org/10.1093/mnras/sty2160}{\emph{Mon. Not. Roy. Astron.
  Soc.} {\bfseries 480} (2018) 5386}
  [\href{https://arxiv.org/abs/1706.04583}{{\ttfamily 1706.04583}}].

\bibitem{Wang:2017htc}
S.~Wang, Y.-F. Wang and D.-M. Xia, \emph{{Constraints on the sum of neutrino
  masses using cosmological data including the latest extended Baryon
  Oscillation Spectroscopic Survey DR14 quasar sample}},
  \href{https://doi.org/10.1088/1674-1137/42/6/065103}{\emph{Chin. Phys.}
  {\bfseries C42} (2018) 065103}
  [\href{https://arxiv.org/abs/1707.00588}{{\ttfamily 1707.00588}}].

\bibitem{Chen:2017ayg}
L.~Chen, Q.-G. Huang and K.~Wang, \emph{{New cosmological constraints with
  extended-Baryon Oscillation Spectroscopic Survey DR14 quasar sample}},
  \href{https://doi.org/10.1140/epjc/s10052-017-5344-1}{\emph{Eur. Phys. J.}
  {\bfseries C77} (2017) 762}
  [\href{https://arxiv.org/abs/1707.02742}{{\ttfamily 1707.02742}}].

\bibitem{Upadhye:2017hdl}
A.~Upadhye, \emph{{Neutrino mass and dark energy constraints from
  redshift-space distortions}},
  \href{https://doi.org/10.1088/1475-7516/2019/05/041}{\emph{JCAP} {\bfseries
  1905} (2019) 041} [\href{https://arxiv.org/abs/1707.09354}{{\ttfamily
  1707.09354}}].

\bibitem{Salvati:2017rsn}
L.~Salvati, M.~Douspis and N.~Aghanim, \emph{{Constraints from thermal
  Sunyaev-Zel’dovich cluster counts and power spectrum combined with CMB}},
  \href{https://doi.org/10.1051/0004-6361/201731990}{\emph{Astron. Astrophys.}
  {\bfseries 614} (2018) A13}
  [\href{https://arxiv.org/abs/1708.00697}{{\ttfamily 1708.00697}}].

\bibitem{Nunes:2017xon}
R.~C. Nunes and A.~Bonilla, \emph{{Probing the properties of relic neutrinos
  using the cosmic microwave background, the Hubble Space Telescope and galaxy
  clusters}}, \href{https://doi.org/10.1093/mnras/stx2661}{\emph{Mon. Not. Roy.
  Astron. Soc.} {\bfseries 473} (2018) 4404}
  [\href{https://arxiv.org/abs/1710.10264}{{\ttfamily 1710.10264}}].

\bibitem{Zennaro:2017qnp}
M.~Zennaro, J.~Bel, J.~Dossett, C.~Carbone and L.~Guzzo, \emph{{Cosmological
  constraints from galaxy clustering in the presence of massive neutrinos}},
  \href{https://doi.org/10.1093/mnras/sty670}{\emph{Mon. Not. Roy. Astron.
  Soc.} {\bfseries 477} (2018) 491}
  [\href{https://arxiv.org/abs/1712.02886}{{\ttfamily 1712.02886}}].

\bibitem{Wang:2018lun}
L.-F. Wang, X.-N. Zhang, J.-F. Zhang and X.~Zhang, \emph{{Impacts of
  gravitational-wave standard siren observation of the Einstein Telescope on
  weighing neutrinos in cosmology}},
  \href{https://doi.org/10.1016/j.physletb.2018.05.027}{\emph{Phys. Lett.}
  {\bfseries B782} (2018) 87}
  [\href{https://arxiv.org/abs/1802.04720}{{\ttfamily 1802.04720}}].

\bibitem{Choudhury:2018adz}
S.~Roy~Choudhury and A.~Naskar, \emph{{Strong Bounds on Sum of Neutrino Masses
  in a 12 Parameter Extended Scenario with Non-Phantom Dynamical Dark Energy
  ($w(z)\geq -1$) with CPL parameterization}},
  \href{https://doi.org/10.1140/epjc/s10052-019-6762-z}{\emph{Eur. Phys. J.}
  {\bfseries C79} (2019) 262}
  [\href{https://arxiv.org/abs/1807.02860}{{\ttfamily 1807.02860}}].

\bibitem{Giusarma:2018jei}
E.~Giusarma, S.~Vagnozzi, S.~Ho, S.~Ferraro, K.~Freese, R.~Kamen-Rubio et~al.,
  \emph{{Scale-dependent galaxy bias, CMB lensing-galaxy cross-correlation, and
  neutrino masses}},
  \href{https://doi.org/10.1103/PhysRevD.98.123526}{\emph{Phys. Rev.}
  {\bfseries D98} (2018) 123526}
  [\href{https://arxiv.org/abs/1802.08694}{{\ttfamily 1802.08694}}].

\bibitem{Loureiro:2018pdz}
A.~Loureiro et~al., \emph{{On The Upper Bound of Neutrino Masses from Combined
  Cosmological Observations and Particle Physics Experiments}},
  \href{https://doi.org/10.1103/PhysRevLett.123.081301}{\emph{Phys. Rev. Lett.}
  {\bfseries 123} (2019) 081301}
  [\href{https://arxiv.org/abs/1811.02578}{{\ttfamily 1811.02578}}].

\bibitem{Hannestad:2016fog}
S.~Hannestad and T.~Schwetz, \emph{{Cosmology and the neutrino mass ordering}},
  \href{https://doi.org/10.1088/1475-7516/2016/11/035}{\emph{JCAP} {\bfseries
  1611} (2016) 035} [\href{https://arxiv.org/abs/1606.04691}{{\ttfamily
  1606.04691}}].

\bibitem{Xu:2016ddc}
L.~Xu and Q.-G. Huang, \emph{{Detecting the Neutrinos Mass Hierarchy from
  Cosmological Data}},
  \href{https://doi.org/10.1007/s11433-017-9125-0}{\emph{Sci. China Phys. Mech.
  Astron.} {\bfseries 61} (2018) 039521}
  [\href{https://arxiv.org/abs/1611.05178}{{\ttfamily 1611.05178}}].

\bibitem{Gerbino:2016ehw}
M.~Gerbino, M.~Lattanzi, O.~Mena and K.~Freese, \emph{{A novel approach to
  quantifying the sensitivity of current and future cosmological datasets to
  the neutrino mass ordering through Bayesian hierarchical modeling}},
  \href{https://doi.org/10.1016/j.physletb.2017.10.052}{\emph{Phys. Lett.}
  {\bfseries B775} (2017) 239}
  [\href{https://arxiv.org/abs/1611.07847}{{\ttfamily 1611.07847}}].

\bibitem{Simpson:2017qvj}
F.~Simpson, R.~Jimenez, C.~Pena-Garay and L.~Verde, \emph{{Strong Bayesian
  Evidence for the Normal Neutrino Hierarchy}},
  \href{https://doi.org/10.1088/1475-7516/2017/06/029}{\emph{JCAP} {\bfseries
  1706} (2017) 029} [\href{https://arxiv.org/abs/1703.03425}{{\ttfamily
  1703.03425}}].

\bibitem{Schwetz:2017fey}
T.~Schwetz, K.~Freese, M.~Gerbino, E.~Giusarma, S.~Hannestad, M.~Lattanzi
  et~al., \emph{{Comment on "Strong Evidence for the Normal Neutrino
  Hierarchy"}},  \href{https://arxiv.org/abs/1703.04585}{{\ttfamily
  1703.04585}}.

\bibitem{Long:2017dru}
A.~J. Long, M.~Raveri, W.~Hu and S.~Dodelson, \emph{{Neutrino Mass Priors for
  Cosmology from Random Matrices}},
  \href{https://doi.org/10.1103/PhysRevD.97.043510}{\emph{Phys. Rev.}
  {\bfseries D97} (2018) 043510}
  [\href{https://arxiv.org/abs/1711.08434}{{\ttfamily 1711.08434}}].

\bibitem{Gariazzo:2018pei}
S.~Gariazzo, M.~Archidiacono, P.~F. de~Salas, O.~Mena, C.~A. Ternes and
  M.~Tórtola, \emph{{Neutrino masses and their ordering: Global Data, Priors
  and Models}},
  \href{https://doi.org/10.1088/1475-7516/2018/03/011}{\emph{JCAP} {\bfseries
  1803} (2018) 011} [\href{https://arxiv.org/abs/1801.04946}{{\ttfamily
  1801.04946}}].

\bibitem{Heavens:2018adv}
A.~F. Heavens and E.~Sellentin, \emph{{Objective Bayesian analysis of neutrino
  masses and hierarchy}},
  \href{https://doi.org/10.1088/1475-7516/2018/04/047}{\emph{JCAP} {\bfseries
  1804} (2018) 047} [\href{https://arxiv.org/abs/1802.09450}{{\ttfamily
  1802.09450}}].

\bibitem{deSalas:2018bym}
P.~F. De~Salas, S.~Gariazzo, O.~Mena, C.~A. Ternes and M.~Tórtola,
  \emph{{Neutrino Mass Ordering from Oscillations and Beyond: 2018 Status and
  Future Prospects}},
  \href{https://doi.org/10.3389/fspas.2018.00036}{\emph{Front. Astron. Space
  Sci.} {\bfseries 5} (2018) 36}
  [\href{https://arxiv.org/abs/1806.11051}{{\ttfamily 1806.11051}}].

\bibitem{Mahony:2019fyb}
C.~Mahony, B.~Leistedt, H.~V. Peiris, J.~Braden, B.~Joachimi, A.~Korn et~al.,
  \emph{{Target Neutrino Mass Precision for Determining the Neutrino
  Hierarchy}}, \href{https://doi.org/10.1103/PhysRevD.101.083513}{\emph{Phys.
  Rev. D} {\bfseries 101} (2020) 083513}
  [\href{https://arxiv.org/abs/1907.04331}{{\ttfamily 1907.04331}}].

\bibitem{Hannestad:2005gj}
S.~Hannestad, \emph{{Neutrino masses and the dark energy equation of state -
  Relaxing the cosmological neutrino mass bound}},
  \href{https://doi.org/10.1103/PhysRevLett.95.221301}{\emph{Phys. Rev. Lett.}
  {\bfseries 95} (2005) 221301}
  [\href{https://arxiv.org/abs/astro-ph/0505551}{{\ttfamily
  astro-ph/0505551}}].

\bibitem{Joudaki:2012fx}
S.~Joudaki, \emph{{Constraints on Neutrino Mass and Light Degrees of Freedom in
  Extended Cosmological Parameter Spaces}},
  \href{https://doi.org/10.1103/PhysRevD.87.083523}{\emph{Phys. Rev.}
  {\bfseries D87} (2013) 083523}
  [\href{https://arxiv.org/abs/1202.0005}{{\ttfamily 1202.0005}}].

\bibitem{Yang:2017amu}
W.~Yang, R.~C. Nunes, S.~Pan and D.~F. Mota, \emph{{Effects of neutrino mass
  hierarchies on dynamical dark energy models}},
  \href{https://doi.org/10.1103/PhysRevD.95.103522}{\emph{Phys. Rev.}
  {\bfseries D95} (2017) 103522}
  [\href{https://arxiv.org/abs/1703.02556}{{\ttfamily 1703.02556}}].

\bibitem{Lorenz:2017fgo}
C.~S. Lorenz, E.~Calabrese and D.~Alonso, \emph{{Distinguishing between
  Neutrinos and time-varying Dark Energy through Cosmic Time}},
  \href{https://doi.org/10.1103/PhysRevD.96.043510}{\emph{Phys. Rev.}
  {\bfseries D96} (2017) 043510}
  [\href{https://arxiv.org/abs/1706.00730}{{\ttfamily 1706.00730}}].

\bibitem{Sutherland:2018ghu}
W.~Sutherland, \emph{{The CMB neutrino mass/vacuum energy degeneracy: a simple
  derivation of the degeneracy slopes}},
  \href{https://doi.org/10.1093/mnras/sty687}{\emph{Mon. Not. Roy. Astron.
  Soc.} {\bfseries 477} (2018) 1913}
  [\href{https://arxiv.org/abs/1803.02298}{{\ttfamily 1803.02298}}].

\bibitem{Sahlen:2018cku}
M.~Sahlén, \emph{{Cluster-Void Degeneracy Breaking: Neutrino Properties and
  Dark Energy}}, \href{https://doi.org/10.1103/PhysRevD.99.063525}{\emph{Phys.
  Rev.} {\bfseries D99} (2019) 063525}
  [\href{https://arxiv.org/abs/1807.02470}{{\ttfamily 1807.02470}}].

\bibitem{DiValentino:2017zyq}
E.~Di~Valentino, A.~Melchiorri, E.~V. Linder and J.~Silk, \emph{{Constraining
  Dark Energy Dynamics in Extended Parameter Space}},
  \href{https://doi.org/10.1103/PhysRevD.96.023523}{\emph{Phys. Rev.}
  {\bfseries D96} (2017) 023523}
  [\href{https://arxiv.org/abs/1704.00762}{{\ttfamily 1704.00762}}].

\bibitem{Vagnozzi:2018jhn}
S.~Vagnozzi, S.~Dhawan, M.~Gerbino, K.~Freese, A.~Goobar and O.~Mena,
  \emph{{Constraints on the sum of the neutrino masses in dynamical dark energy
  models with $w(z) \geq -1$ are tighter than those obtained in $\Lambda$CDM}},
  \href{https://doi.org/10.1103/PhysRevD.98.083501}{\emph{Phys. Rev.}
  {\bfseries D98} (2018) 083501}
  [\href{https://arxiv.org/abs/1801.08553}{{\ttfamily 1801.08553}}].

\bibitem{Zhang:2015uhk}
X.~Zhang, \emph{{Impacts of dark energy on weighing neutrinos after Planck
  2015}}, \href{https://doi.org/10.1103/PhysRevD.93.083011}{\emph{Phys. Rev.}
  {\bfseries D93} (2016) 083011}
  [\href{https://arxiv.org/abs/1511.02651}{{\ttfamily 1511.02651}}].

\bibitem{Wang:2016tsz}
S.~Wang, Y.-F. Wang, D.-M. Xia and X.~Zhang, \emph{{Impacts of dark energy on
  weighing neutrinos: mass hierarchies considered}},
  \href{https://doi.org/10.1103/PhysRevD.94.083519}{\emph{Phys. Rev.}
  {\bfseries D94} (2016) 083519}
  [\href{https://arxiv.org/abs/1608.00672}{{\ttfamily 1608.00672}}].

\bibitem{Archidiacono:2020dvx}
M.~Archidiacono, S.~Hannestad and J.~Lesgourgues, \emph{{What will it take to
  measure individual neutrino mass states using cosmology?}},
  \href{https://arxiv.org/abs/2003.03354}{{\ttfamily 2003.03354}}.

\bibitem{DiValentino:2016foa}
{\scshape CORE} collaboration, E.~Di~Valentino et~al., \emph{{Exploring cosmic
  origins with CORE: Cosmological parameters}},
  \href{https://doi.org/10.1088/1475-7516/2018/04/017}{\emph{JCAP} {\bfseries
  04} (2018) 017} [\href{https://arxiv.org/abs/1612.00021}{{\ttfamily
  1612.00021}}].

\bibitem{Handley:2018gel}
W.~Handley and M.~Millea, \emph{{Maximum-Entropy Priors with Derived Parameters
  in a Specified Distribution}},
  \href{https://doi.org/10.3390/e21030272}{\emph{Entropy} {\bfseries 21} (2019)
  272} [\href{https://arxiv.org/abs/1804.08143}{{\ttfamily 1804.08143}}].

\bibitem{Hu:2007pj}
W.~Hu and I.~Sawicki, \emph{{A Parameterized Post-Friedmann Framework for
  Modified Gravity}},
  \href{https://doi.org/10.1103/PhysRevD.76.104043}{\emph{Phys. Rev.}
  {\bfseries D76} (2007) 104043}
  [\href{https://arxiv.org/abs/0708.1190}{{\ttfamily 0708.1190}}].

\bibitem{Chevallier:2000qy}
M.~Chevallier and D.~Polarski, \emph{{Accelerating universes with scaling dark
  matter}}, \href{https://doi.org/10.1142/S0218271801000822}{\emph{Int. J. Mod.
  Phys.} {\bfseries D10} (2001) 213}
  [\href{https://arxiv.org/abs/gr-qc/0009008}{{\ttfamily gr-qc/0009008}}].

\bibitem{Linder:2002et}
E.~V. Linder, \emph{{Exploring the expansion history of the universe}},
  \href{https://doi.org/10.1103/PhysRevLett.90.091301}{\emph{Phys. Rev. Lett.}
  {\bfseries 90} (2003) 091301}
  [\href{https://arxiv.org/abs/astro-ph/0208512}{{\ttfamily
  astro-ph/0208512}}].

\bibitem{Calabrese:2008rt}
E.~Calabrese, A.~Slosar, A.~Melchiorri, G.~F. Smoot and O.~Zahn, \emph{{Cosmic
  Microwave Weak lensing data as a test for the dark universe}},
  \href{https://doi.org/10.1103/PhysRevD.77.123531}{\emph{Phys. Rev.}
  {\bfseries D77} (2008) 123531}
  [\href{https://arxiv.org/abs/0803.2309}{{\ttfamily 0803.2309}}].

\bibitem{Matsumura:2013aja}
T.~Matsumura et~al., \emph{{Mission design of LiteBIRD}},
  \href{https://arxiv.org/abs/1311.2847}{{\ttfamily 1311.2847}}.

\bibitem{Renzi:2017cbg}
F.~Renzi, E.~Di~Valentino and A.~Melchiorri, \emph{{Cornering the Planck
  $A_{lens}$ anomaly with future CMB data}},
  \href{https://doi.org/10.1103/PhysRevD.97.123534}{\emph{Phys. Rev.}
  {\bfseries D97} (2018) 123534}
  [\href{https://arxiv.org/abs/1712.08758}{{\ttfamily 1712.08758}}].

\bibitem{Lewis:2002ah}
A.~Lewis and S.~Bridle, \emph{{Cosmological parameters from CMB and other data:
  A Monte Carlo approach}},
  \href{https://doi.org/10.1103/PhysRevD.66.103511}{\emph{Phys. Rev.}
  {\bfseries D66} (2002) 103511}
  [\href{https://arxiv.org/abs/astro-ph/0205436}{{\ttfamily
  astro-ph/0205436}}].

\bibitem{Lewis:1999bs}
A.~Lewis, A.~Challinor and A.~Lasenby, \emph{{Efficient computation of CMB
  anisotropies in closed FRW models}},
  \href{https://doi.org/10.1086/309179}{\emph{Astrophys. J.} {\bfseries 538}
  (2000) 473} [\href{https://arxiv.org/abs/astro-ph/9911177}{{\ttfamily
  astro-ph/9911177}}].

\bibitem{doi:10.1080/10618600.1998.10474787}
S.~P. Brooks and A.~Gelman, \emph{General Methods for Monitoring Convergence of
  Iterative Simulations},
  \href{https://doi.org/10.1080/10618600.1998.10474787}{\emph{Journal of
  Computational and Graphical Statistics} {\bfseries 7} (1998) 434}
  [\href{https://arxiv.org/abs/https://www.tandfonline.com/doi/pdf/10.1080/10618600.1998.10474787}{{\ttfamily
  https://www.tandfonline.com/doi/pdf/10.1080/10618600.1998.10474787}}].

\bibitem{Handley}
W.~Handley, \emph{{CosmoChord: Planck 2018 update}},
  \href{https://arxiv.org/abs/https://doi.org/10.5281/zenodo.3370086}{{\ttfamily
  https://doi.org/10.5281/zenodo.3370086}}.

\bibitem{Handley:2015fda}
W.~J. Handley, M.~P. Hobson and A.~N. Lasenby, \emph{{PolyChord: nested
  sampling for cosmology}},
  \href{https://doi.org/10.1093/mnrasl/slv047}{\emph{Mon. Not. Roy. Astron.
  Soc.} {\bfseries 450} (2015) L61}
  [\href{https://arxiv.org/abs/1502.01856}{{\ttfamily 1502.01856}}].

\bibitem{10.1093/mnras/stv1911}
W.~J. Handley, M.~P. Hobson and A.~N. Lasenby, \emph{{polychord:
  next-generation nested sampling}},
  \href{https://doi.org/10.1093/mnras/stv1911}{\emph{Monthly Notices of the
  Royal Astronomical Society} {\bfseries 453} (2015) 4384}
  [\href{https://arxiv.org/abs/1506.00171}{{\ttfamily 1506.00171}}].

\bibitem{Aghanim:2019ame}
{\scshape Planck} collaboration, N.~Aghanim et~al., \emph{{Planck 2018 results.
  V. CMB power spectra and likelihoods}},
  \href{https://arxiv.org/abs/1907.12875}{{\ttfamily 1907.12875}}.

\bibitem{Ade:2018gkx}
{\scshape BICEP2, Keck Array} collaboration, P.~A.~R. Ade et~al., \emph{{BICEP2
  / Keck Array x: Constraints on Primordial Gravitational Waves using Planck,
  WMAP, and New BICEP2/Keck Observations through the 2015 Season}},
  \href{https://doi.org/10.1103/PhysRevLett.121.221301}{\emph{Phys. Rev. Lett.}
  {\bfseries 121} (2018) 221301}
  [\href{https://arxiv.org/abs/1810.05216}{{\ttfamily 1810.05216}}].

\bibitem{Alam:2016hwk}
{\scshape BOSS} collaboration, S.~Alam et~al., \emph{{The clustering of
  galaxies in the completed SDSS-III Baryon Oscillation Spectroscopic Survey:
  cosmological analysis of the DR12 galaxy sample}},
  \href{https://doi.org/10.1093/mnras/stx721}{\emph{Mon. Not. Roy. Astron.
  Soc.} {\bfseries 470} (2017) 2617}
  [\href{https://arxiv.org/abs/1607.03155}{{\ttfamily 1607.03155}}].

\bibitem{Ross:2014qpa}
A.~J. Ross, L.~Samushia, C.~Howlett, W.~J. Percival, A.~Burden and M.~Manera,
  \emph{{The clustering of the SDSS DR7 main Galaxy sample – I. A 4 per cent
  distance measure at $z = 0.15$}},
  \href{https://doi.org/10.1093/mnras/stv154}{\emph{Mon. Not. Roy. Astron.
  Soc.} {\bfseries 449} (2015) 835}
  [\href{https://arxiv.org/abs/1409.3242}{{\ttfamily 1409.3242}}].

\bibitem{Beutler:2011hx}
F.~Beutler, C.~Blake, M.~Colless, D.~H. Jones, L.~Staveley-Smith, L.~Campbell
  et~al., \emph{{The 6dF Galaxy Survey: Baryon Acoustic Oscillations and the
  Local Hubble Constant}},
  \href{https://doi.org/10.1111/j.1365-2966.2011.19250.x}{\emph{Mon. Not. Roy.
  Astron. Soc.} {\bfseries 416} (2011) 3017}
  [\href{https://arxiv.org/abs/1106.3366}{{\ttfamily 1106.3366}}].

\bibitem{Scolnic:2017caz}
D.~M. Scolnic et~al., \emph{{The Complete Light-curve Sample of
  Spectroscopically Confirmed SNe Ia from Pan-STARRS1 and Cosmological
  Constraints from the Combined Pantheon Sample}},
  \href{https://doi.org/10.3847/1538-4357/aab9bb}{\emph{Astrophys. J.}
  {\bfseries 859} (2018) 101}
  [\href{https://arxiv.org/abs/1710.00845}{{\ttfamily 1710.00845}}].

\bibitem{Ade:2015xua}
{\scshape Planck} collaboration, P.~A.~R. Ade et~al., \emph{{Planck 2015
  results. XIII. Cosmological parameters}},
  \href{https://doi.org/10.1051/0004-6361/201525830}{\emph{Astron. Astrophys.}
  {\bfseries 594} (2016) A13}
  [\href{https://arxiv.org/abs/1502.01589}{{\ttfamily 1502.01589}}].

\bibitem{Reichardt2016}
C.~L. Reichardt, \emph{Understanding the Epoch of Cosmic Reionization:
  Challenges and Progress}, pp.~227--245.
\newblock Springer International Publishing, Cham, Switzerland, 2016.

\bibitem{Hou:2012xq}
Z.~Hou et~al., \emph{{Constraints on Cosmology from the Cosmic Microwave
  Background Power Spectrum of the 2500 deg$^2$ SPT-SZ Survey}},
  \href{https://doi.org/10.1088/0004-637X/782/2/74}{\emph{Astrophys. J.}
  {\bfseries 782} (2014) 74} [\href{https://arxiv.org/abs/1212.6267}{{\ttfamily
  1212.6267}}].

\bibitem{Kitching:2016hvn}
T.~D. Kitching, L.~Verde, A.~F. Heavens and R.~Jimenez, \emph{{Discrepancies
  between CFHTLenS cosmic shear and Planck: new physics or systematic
  effects?}}, \href{https://doi.org/10.1093/mnras/stw707}{\emph{Mon. Not. Roy.
  Astron. Soc.} {\bfseries 459} (2016) 971}
  [\href{https://arxiv.org/abs/1602.02960}{{\ttfamily 1602.02960}}].

\bibitem{1100705}
H.~{Akaike}, \emph{A new look at the statistical model identification},
  \href{https://doi.org/10.1109/TAC.1974.1100705}{\emph{IEEE Transactions on
  Automatic Control} {\bfseries 19} (1974) 716}.

\bibitem{Hamann:2012fe}
J.~Hamann, S.~Hannestad and Y.~Y.~Y. Wong, \emph{{Measuring neutrino masses
  with a future galaxy survey}},
  \href{https://doi.org/10.1088/1475-7516/2012/11/052}{\emph{JCAP} {\bfseries
  1211} (2012) 052} [\href{https://arxiv.org/abs/1209.1043}{{\ttfamily
  1209.1043}}].

\bibitem{Brinckmann:2018owf}
T.~Brinckmann, D.~C. Hooper, M.~Archidiacono, J.~Lesgourgues and T.~Sprenger,
  \emph{{The promising future of a robust cosmological neutrino mass
  measurement}},
  \href{https://doi.org/10.1088/1475-7516/2019/01/059}{\emph{JCAP} {\bfseries
  1901} (2019) 059} [\href{https://arxiv.org/abs/1808.05955}{{\ttfamily
  1808.05955}}].

\bibitem{Sprenger:2018tdb}
T.~Sprenger, M.~Archidiacono, T.~Brinckmann, S.~Clesse and J.~Lesgourgues,
  \emph{{Cosmology in the era of Euclid and the Square Kilometre Array}},
  \href{https://doi.org/10.1088/1475-7516/2019/02/047}{\emph{JCAP} {\bfseries
  1902} (2019) 047} [\href{https://arxiv.org/abs/1801.08331}{{\ttfamily
  1801.08331}}].

\bibitem{Chudaykin:2019ock}
A.~Chudaykin and M.~M. Ivanov, \emph{{Measuring neutrino masses with
  large-scale structure: Euclid forecast with controlled theoretical error}},
  \href{https://doi.org/10.1088/1475-7516/2019/11/034}{\emph{JCAP} {\bfseries
  1911} (2019) 034} [\href{https://arxiv.org/abs/1907.06666}{{\ttfamily
  1907.06666}}].

\bibitem{Choudhury:2018sbz}
S.~Roy~Choudhury and S.~Choubey, \emph{{Constraining light sterile neutrino
  mass with the BICEP2/Keck Array 2014 B-mode polarization data}},
  \href{https://doi.org/10.1140/epjc/s10052-019-7063-2}{\emph{Eur. Phys. J.}
  {\bfseries C79} (2019) 557}
  [\href{https://arxiv.org/abs/1807.10294}{{\ttfamily 1807.10294}}].

\bibitem{Array:2016afx}
{\scshape BICEP2, Keck Array} collaboration, P.~A.~R. Ade et~al., \emph{{BICEP2
  / Keck Array VIII: Measurement of gravitational lensing from large-scale
  B-mode polarization}},
  \href{https://doi.org/10.3847/1538-4357/833/2/228}{\emph{Astrophys. J.}
  {\bfseries 833} (2016) 228}
  [\href{https://arxiv.org/abs/1606.01968}{{\ttfamily 1606.01968}}].

\bibitem{Linder:2007wa}
E.~V. Linder, \emph{{The Dynamics of Quintessence, The Quintessence of
  Dynamics}}, \href{https://doi.org/10.1007/s10714-007-0550-z}{\emph{Gen. Rel.
  Grav.} {\bfseries 40} (2008) 329}
  [\href{https://arxiv.org/abs/0704.2064}{{\ttfamily 0704.2064}}].

\bibitem{Erben:2012zw}
T.~Erben et~al., \emph{{CFHTLenS: The Canada-France-Hawaii Telescope Lensing
  Survey - Imaging Data and Catalogue Products}},
  \href{https://doi.org/10.1093/mnras/stt928}{\emph{Mon. Not. Roy. Astron.
  Soc.} {\bfseries 433} (2013) 2545}
  [\href{https://arxiv.org/abs/1210.8156}{{\ttfamily 1210.8156}}].

\bibitem{Hildebrandt:2016iqg}
H.~Hildebrandt et~al., \emph{{KiDS-450: Cosmological parameter constraints from
  tomographic weak gravitational lensing}},
  \href{https://doi.org/10.1093/mnras/stw2805}{\emph{Mon. Not. Roy. Astron.
  Soc.} {\bfseries 465} (2017) 1454}
  [\href{https://arxiv.org/abs/1606.05338}{{\ttfamily 1606.05338}}].

\bibitem{Howlett:2012mh}
C.~Howlett, A.~Lewis, A.~Hall and A.~Challinor, \emph{{CMB power spectrum
  parameter degeneracies in the era of precision cosmology}},
  \href{https://doi.org/10.1088/1475-7516/2012/04/027}{\emph{JCAP} {\bfseries
  1204} (2012) 027} [\href{https://arxiv.org/abs/1201.3654}{{\ttfamily
  1201.3654}}].

\bibitem{Blennow:2013kga}
M.~Blennow, \emph{{On the Bayesian approach to neutrino mass ordering}},
  \href{https://doi.org/10.1007/JHEP01(2014)139}{\emph{JHEP} {\bfseries 01}
  (2014) 139} [\href{https://arxiv.org/abs/1311.3183}{{\ttfamily 1311.3183}}].

\bibitem{Hall:2012kg}
A.~C. Hall and A.~Challinor, \emph{{Probing the neutrino mass hierarchy with
  CMB weak lensing}},
  \href{https://doi.org/10.1111/j.1365-2966.2012.21493.x}{\emph{Mon. Not. Roy.
  Astron. Soc.} {\bfseries 425} (2012) 1170}
  [\href{https://arxiv.org/abs/1205.6172}{{\ttfamily 1205.6172}}].

\bibitem{Jeffreys}
H.~Jeffreys, \emph{{Theory of Probability}}. The International series of
  monographs on physics. Clarendon Press, 1961.

\bibitem{DiValentino:2019dzu}
E.~Di~Valentino, A.~Melchiorri and J.~Silk, \emph{{Cosmological constraints in
  extended parameter space from the Planck 2018 Legacy release}},
  \href{https://doi.org/10.1088/1475-7516/2020/01/013}{\emph{JCAP} {\bfseries
  2001} (2020) 013} [\href{https://arxiv.org/abs/1908.01391}{{\ttfamily
  1908.01391}}].

\bibitem{Acciarri:2015uup}
{\scshape DUNE} collaboration, R.~Acciarri et~al., \emph{{Long-Baseline
  Neutrino Facility (LBNF) and Deep Underground Neutrino Experiment (DUNE)}:
  {Conceptual Design Report, Volume 2: The Physics Program for DUNE at LBNF}},
  \href{https://arxiv.org/abs/1512.06148}{{\ttfamily 1512.06148}}.

\bibitem{An:2015jdp}
{\scshape JUNO} collaboration, F.~An et~al., \emph{{Neutrino Physics with
  JUNO}}, \href{https://doi.org/10.1088/0954-3899/43/3/030401}{\emph{J. Phys.
  G} {\bfseries 43} (2016) 030401}
  [\href{https://arxiv.org/abs/1507.05613}{{\ttfamily 1507.05613}}].

\end{thebibliography}\endgroup
\bibliographystyle{JHEP}

\end{document}